\documentclass[hyper,11pt,letterpaper]{JHEP3}
\usepackage[dvips]{epsfig}
\usepackage{amsmath,amssymb,epsf}
\usepackage{cite,overpic,epic}
\usepackage{graphicx}
\usepackage{dcolumn}
\usepackage{bm}

\def\coeff#1#2{{\textstyle {\frac {#1}{#2}}}}
\def\half{\coeff 12}

\newcommand{\be}{\begin{equation}}
\newcommand{\ee}{\end{equation}}
\newcommand{\bea}{\begin{eqnarray}}
\newcommand{\eea}{\end{eqnarray}}
\newcommand{\ba}{\begin{eqnarray}}
\newcommand{\ea}{\end{eqnarray}}

\newcommand{\beq}{\begin{equation}}
\newcommand{\eeq}{\end{equation}}
\newcommand{\beqa}{\begin{eqnarray}}
\newcommand{\eeqa}{\end{eqnarray}}
\newcommand{\beqar}{\begin{eqnarray*}}
\newcommand{\eeqar}{\end{eqnarray*}}

\newcommand{\N}{{\mathcal{N}}}
\newcommand{\R}{{\mathbb{R}}}
\renewcommand{\Re}{{\rm Re}\,}
\renewcommand{\Im}{{\rm Im}\,}

\def\a{\alpha}
\def\b{\beta}

\def\d{\delta}
\def\e{\epsilon}           
  
\def\g{\gamma}

\def\ps{\psi}
\def\lam{\lambda}
\def\m{\mu}

\def\o{\omega}  
\def\p{\pi}                
\def\th{\theta}                   
\def\r{\rho}                                     
\def\t{\tau}

\def\D{\Delta}

\def\Nc{N_c}
\def\Nf{N_f}
\def\Dp{D$p$}
\def\Nfour{\mathcal N\,{=}\,4}
\def\Ntwo{\mathcal N\,{=}\,2}

\def\Om{{\cal{O}}_m}
\def\lang{\langle}
\def\rang{\rangle}
\def\ra{\rightarrow}

\def\half{{\textstyle \frac 12}}

\title{\LARGE Critical Exponents from AdS/CFT with Flavor}

\author{Andreas Karch,\!$^1$\footnotemark[1]\,
Andy O'Bannon,\!$^{1,2}$\footnotemark[2]\,
and Laurence G. Yaffe\!$^1$\footnotemark[3]
\\
$^1$Department of Physics, University of Washington, \\ 3910 15th Ave. NE, Seattle, WA 98195-1560, U.S.A.
\\
$^2$Max Planck Institut f\"{u}r Physik (Werner Heisenberg Institut) \\ F\"{o}hringer Ring 6, 80805 M\"{u}nchen, Germany}

\footnotetext[1]{E-mail: \email{karch@phys.washington.edu}}
\footnotetext[2]{E-mail: \email{ahob@mppmu.mpg.de}}
\footnotetext[3]{E-mail: \email{yaffe@phys.washington.edu}}

\abstract
{
We use the AdS/CFT correspondence to study the thermodynamics of massive $\Ntwo$ supersymmetric hypermultiplet flavor fields coupled to
$\Nfour$ supersymmetric $SU(N_c)$ Yang-Mills theory, formulated on curved four-manifolds,
in the limits of large $\Nc$ and large 't Hooft coupling.
The gravitational duals are
probe D-branes in global thermal $AdS$. These D-branes may undergo a
topology-changing transition in the bulk. The D-brane embeddings near
the point of the topology change exhibit a scaling symmetry.
The associated scaling exponents can be either
real- or complex-valued.
Which regime applies depends on the dimensionality of a collapsing
submanifold in the critical embedding.
When the scaling exponents are complex-valued, a
first-order transition associated with the flavor fields appears in the
dual field theory.
Real scaling exponents are expected to be associated with a
continuous transition in the dual field theory.
For one example with real exponents, the D7-brane, we
study the transition in detail. We find two field theory observables that
diverge at the critical point, and we compute the associated critical
exponents. We also present analytic and numerical evidence that the
transition expresses itself in the meson spectrum as a non-analyticity
at the critical point. We argue that the transition we study is a true
phase transition only when the 't Hooft coupling is strictly infinite.
}


\preprint{MPP-2009-70}
\begin{document}

\section{Introduction}  \label{intro}

The Anti-de Sitter/Conformal Field Theory ($AdS$/CFT) correspondence is the conjectured equivalence between $\Nfour$ supersymmetric $SU(N_c)$ Yang-Mills (SYM) theory and type IIB string theory formulated on the background spacetime
$AdS_5 \times S^5$ \cite{Maldacena:1997re,Witten:1998qj,Gubser:1998bc}. Here, $AdS_5$ is a ``Poincar\'e patch'' of five-dimensional anti-de Sitter space, and $S^5$ is a five-sphere. The $AdS_5 \times S^5$ spacetime arises as the near-horizon geometry of $\Nc \ra \infty$ coincident D3-branes. Taking large $\Nc$, small string coupling $g_s$, and fixed but large $g_s \Nc$, the low-energy dynamics of the string theory is well-approximated by type IIB supergravity. In the SYM theory these correspond to the limits of large $\Nc$ and small $g_{YM}^2$, but large 't Hooft coupling $\lambda \equiv g_{YM}^2 \Nc$.

Thermal equilibrium of the
$\Nfour$ SYM theory corresponds to non-extremal D3-branes, whose
near-horizon geometry is five-dimensional AdS-Schwarzschild times
$S^5$ \cite{Witten:1998zw,Gubser:1996de}. The temperature of the
$\Nfour$ SYM theory coincides with the Hawking temperature of the
AdS-Schwarzschild black hole.

The $\Nfour$ SYM theory only contains fields in the adjoint
representation of the gauge group. To make the theory more closely
resemble the theory of strong interactions, Quantum Chromodynamics
(QCD), we may introduce massive ``flavor'' fields in the fundamental
representation of the gauge group. We will introduce a finite number $\Nf$
of them, so that $\Nf \ll \Nc$ in the large $\Nc$ limit, and work to leading
non-trivial order in $\Nf / \Nc$. These flavor fields appear in the supergravity
description as $\Nf$ probe D-branes \cite{Karch:2002sh}, or ``flavor
branes.''
In particular, a probe D7-brane describes a massive $\Ntwo$
supersymmetric hypermultiplet propagating in 3+1 dimensions
and interacting with the $\Nfour$ SYM fields.
In the $\Nf \ll \Nc$ limit, the contribution that these  flavor
branes make to the stress-energy tensor is dwarfed by the contribution
of the $\Nc$ D3-branes. To leading order in $\Nf / \Nc$, one may
neglect the back-reaction of the flavor branes on the geometry, hence the
label ``probe.'' The flavor brane action is then the Dirac-Born-Infeld
(DBI) action characterizing the embedding of the D-brane in the
background spacetime.

In the supergravity description, the embedding of the D7-brane undergoes
a topology-changing transition as the temperature increases.
This corresponds in the dual field
theory to a first-order phase transition associated with the flavor fields
\cite{Babington:2003vm,Kirsch:2004km,Ghoroku:2005tf,Apreda:2005yz,Mateos:2006nu,Albash:2006ew,Karch:2006bv,Mateos:2007vn}.
In Refs.~\cite{Mateos:2006nu,Mateos:2007vn}, analysis of a general
class of probe D-brane systems (based on supersymmetry-preserving D$p$-D$q$ intersections) revealed that in these systems the transitions associated with
probe D-branes are generically first order.

In Ref.~\cite{Karch:2006bv}, two of the present authors analyzed the
thermodynamics of flavor fields in the SYM theory formulated on a
3-sphere. The supergravity dual in this case does not arise as the
near-horizon limit of any known D-brane construction in string theory:
$\Nfour$ SYM formulated on a 3-sphere, in the 't Hooft limit and at
large 't Hooft coupling, is holographically dual to supergravity
formulated on \textit{global} $AdS$, as opposed to Poincar\'{e}
patch $AdS$. From the SYM theory perspective, the 3-sphere
introduces a scale into the CFT allowing for a thermal phase
transition. (Without such a scale to set a transition temperature,
there can be no phase transition at any non-zero temperature.)
At high temperature, relative to the
inverse 3-sphere radius, the $\Nfour$ SYM theory is in a
``deconfined'' phase, in which the free energy scales as $N_c^2$. At low temperature, however, the free energy is order one, {\em i.e.}, $O(N_c^0)$. A first-order phase transition separates
the two phases. In the supergravity description, the transition
appears as a Hawking-Page transition
\cite{Hawking:1982dh,Witten:1998zw} in which the high-temperature
phase is global $AdS$-Schwarzschild and the low-temperature phase is
global $AdS$ with a periodic time direction, which we will call global
thermal $AdS$.

We may introduce a probe D7-brane into this background,
which now corresponds to adding a flavor
hypermultiplet to $\Nfour$ SYM defined on $S^3$.
In the high-temperature phase,
a probe D7-brane undergoes a
topology-changing transition that is essentially the same as that in
Poincar\'{e} patch $AdS$-Schwarzschild
\cite{Babington:2003vm,Kirsch:2004km,Ghoroku:2005tf,Apreda:2005yz,Mateos:2006nu,Albash:2006ew,Karch:2006bv,Mateos:2007vn}.
This, once again, represents a
first-order transition in the flavor physics of the dual field theory \cite{Karch:2006bv}. In the low-temperature phase the probe D7-brane undergoes a different topology-changing transition. The surprising result of Ref.~\cite{Karch:2006bv} was that this transition does not appear to correspond to a first-order transition in the SYM theory.

Our goal in this paper is to understand the nature of this transition. To
do so, we will provide a more general context in which to understand the D7-brane's transition. We will examine other probe D-branes that are supersymmetric at zero temperature (and in the Poincar\'{e} patch), in particular a probe D5-brane \cite{Karch:2000gx,DeWolfe:2001pq,Erdmenger:2002ex}, which describes flavor fields propagating in 2+1 dimensions. We will also study the $\Nfour$ SYM theory formulated on certain other curved manifolds.

As in Refs.~\cite{Mateos:2006nu,Mateos:2007vn,Filev:2007qu,Filev:2008xt},
an essential tool in our analysis will be
scaling symmetries of probe D-brane embeddings.
The crucial ingredient for the arguments
of Refs.~\cite{Mateos:2006nu,Mateos:2007vn,Filev:2007qu,Filev:2008xt} was that
certain scaling exponents associated with probe D-brane embedding
solutions were \textit{complex-valued}, which implied
the existence of infinitely many solutions of the embedding equations
(and boundary conditions). Such ``multi-valued'' supergravity solutions
imply the presence of discontinuities in SYM theory observables.
As external parameters, such as the hypermultiplet mass, are varied,
the physically relevant solution can jump from one branch to another,
leading to discontinuities in generic observables. (The details of the argument will be reviewed below.)

We will find that the D7-brane in global thermal $AdS$ has
\textit{real-valued} exponents, and that this will lead to very different
physical consequences. Our general analysis will reveal the condition
that determines whether the scaling exponents are complex or real. In
the D-brane's topology change, we can identify a ``critical solution''
intermediate between the two topologies. The dimension of the submanifold
that collapses to zero volume in this critical solution determines
whether the scaling exponents will be real or complex.

For the D7-brane in global thermal $AdS$ we will further characterize the transition by computing two observables that are especially convenient to evaluate using the supergravity description: the expectation value of the hypermultiplet mass operator ({\em i.e.}, the supersymmetric completion of $\bar\psi\psi$), and the expectation value of a particular supersymmetric Polyakov loop correlator. We will find that both have derivatives that diverge near the critical point, and we will compute the associated critical exponents. We will also compute the meson spectrum of the SYM theory from fluctuations of the D7-brane's worldvolume fields. We will find that the transition appears in the meson spectrum as a non-analyticity, a cusp, in the spectrum of one scalar meson. Lastly, we will argue that the transition is only a genuine phase transition in the flavor physics at infinite 't Hooft coupling, that is, the divergences we find in derivatives of the free energy only exist when $\lambda$ is strictly infinite.

This paper is organized as follows. In section \ref{systems} we present in greater detail the systems that we will study. In section \ref{scaling}, we perform the scaling analysis, compute the scaling exponents, and present numerical evidence that complex exponents signal a first-order transition in the SYM theory. In section \ref{transition}, we compute the critical exponents and meson spectrum for the SYM theory dual to the probe D7-brane in global thermal $AdS$, and argue that
the transition will be smoothed out at finite $\lambda$. We conclude with a brief discussion in section \ref{conclusion}.

\section{The Systems in Question}  \label{systems}

We begin with a more precise description of the supergravity systems we will study. After reviewing global $AdS$ and introducing our conventions, we describe the relevant probe D-brane embeddings in greater detail.

\subsection{The Background Geometries}  \label{background}

We are interested in $\Nfour$ SYM theory formulated on curved spatial sections. We work in the limits of large $N_c$ and large 't Hooft coupling, $\lambda \equiv g_{YM}^2 N_c \gg 1$ so that, via the AdS/CFT correspondence, we may compute observables in the SYM theory using supergravity. We are thus interested in global $AdS_5$, whose boundary is a curved four-manifold. Global $AdS_5$ does not arise from any known string theory construction (such as a near-horizon geometry of very many D-branes), so we must assume that the conjectured equivalence of theories, originally motivated by the black
D3-brane solution of type IIB string theory \cite{Maldacena:1997re}, can be extended to include supergravity on global $AdS_5$.

As we will be studying the thermodynamics of the SYM theory, we introduce a temperature $T$ in the standard fashion by working in Euclidean signature and compactifying the time direction into a circle of radius $R_1 = 1 / (2 \pi T)$. We will mainly be interested in the SYM theory formulated on a spatial 3-sphere (times time). We denote the radius of the 3-sphere as $R_3$. We will thus be studying the SYM theory formulated on boundary spacetimes such as $S^1 \times S^3$.

We will use a global thermal $AdS_5$ metric
\begin{equation}
\label{thermaladsmetric}
ds^2 = d\r^2 + \cosh^{2}\!\r \; d\t^2 + \sinh^2\!\r \; d\Omega^{2}_{3} \,,
\end{equation}
where we have set the curvature radius of $AdS_5$ equal to one. We will work in these units throughout. In these units, we convert between string theory and SYM quantities using the relation $\alpha'^{-2} = 4 \pi g_s N_c = g_{YM}^2 N_c = \lambda$, where $\alpha'$ is the square of the string length: $\alpha' \equiv \ell_s^2$. In Eq. (\ref{thermaladsmetric}), $\r$ is the radial coordinate, $\t$ is the compact Euclidean time coordinate of period $1/T$, and $d\Omega^{2}_{3}$ is the metric of a unit 3-sphere. The center of the $AdS$ space is at $\r = 0$, and the boundary is at $\r = \infty$. Notice in particular that the 3-sphere collapses to zero volume at the center of $AdS$.

We will also frequently find another coordinate system convenient, a so-called Fefferman-Graham coordinate system \cite{Fefferman} in
which the radial coordinate is $z = e^{-\rho}$, so that the center of
$AdS_5$ is at $z = 1$ and the boundary is at $z=0$.
The global $AdS_5$-Schwarzschild metric, written in this Fefferman-Graham
coordinate system, is
\begin{subequations}
\begin{equation}
\label{adsbha}
ds^2 = \frac 1{z^2} \left[ dz^2 + g_{ij} \, dx^i dx^j \right],
\end{equation}
with
\begin{align}
\label{adsbhb}
g_{ij} \, dx^i dx^j &\equiv
\frac{1}{4}\, (1-z^4 / z_H^4)^2 \; {\mathcal{F}}(z)^{-1} \; d\t^2
+ \frac{1}{4} \, {\mathcal{F}}(z) \; d\Omega^{2}_{3} \,,
\\
\label{adsbhc}
{\mathcal{F}}(z) &= 1 - 2z^2 + z^4/z_H^4 \,,
\end{align}
\end{subequations}
and $z_H \equiv [1+4(\pi T)^4]^{-1/4}$ is the horizon radius. Setting $z_H = 1$ produces the global thermal $AdS_5$ metric, Eq.~(\ref{thermaladsmetric}), in Fefferman-Graham coordinates. Notice that in our units the radius of the 3-sphere at the boundary is $R_3 = 1/2$.

When discussing global thermal $AdS_5$, we will use the $\rho$ coordinate unless stated otherwise. When discussing $AdS_5$-Schwarzschild, we will use only the $z$ coordinate.

A Hawking-Page transition \cite{Hawking:1982dh} connects global thermal $AdS_5$ and global $AdS_5$-Schwarz\-schild.
This is a first order transition associated with black hole condensation. The two metrics written above are both solutions to Einstein's equation with constant negative curvature and an asymptotic boundary that is $S^1 \times S^3$.
The difference of the Einstein-Hilbert action for these two solutions may be interpreted as a difference in free energies
(divided by the temperature), and thus determines which geometry is thermodynamically preferred. Thermal $AdS_5$ is preferred (has lower free energy) at low temperatures, but a first-order transition occurs at a non-zero critical
temperature $T_{HP} = \frac{3}{2 \pi}$ (in units of the $AdS$ radius), above which $AdS_5$-Schwarzschild is preferred \cite{Hawking:1982dh,Witten:1998zw}. This is a topology-changing transition: the topology of thermal $AdS_5$ is $\R^4 \times S^1$ while that of $AdS_5$-Schwarzschild is $\R^2 \times S^3$ \cite{Hawking:1982dh,Witten:1998zw}.

This transition is interpreted in the boundary SYM theory as a deconfinement transition \cite{Witten:1998zw}. As the SYM theory is in finite volume, we must be careful to define what we mean by the word ``confinement.'' A sufficient definition for our purposes may be given in terms of the large-$\Nc$ behavior of the free energy: if the order $\Nc^2$ contribution to the free energy is zero, we declare the theory to be in a confined phase, whereas if the order $\Nc^2$
contribution is nonzero we declare the theory to be in a deconfined phase.
In this sense, the $\Nfour$ theory on $S^1 \times S^3$ is in a confined phase at
low temperature and in a deconfined phase at high temperature, with
a first-order transition separating the two phases.
In the absence of fundamental representation matter fields,
this criterion is equivalent to defining a deconfined phase as one in which the
$\mathbb Z_{\Nc}$ center symmetry is spontaneously broken
in the $\Nc\to\infty$ limit,
which is possible even in finite spatial volume.

Notice also that the only scales in the SYM theory are the temperature $T$ and the radius of the 3-sphere $R_3$, and hence the only physically meaningful quantity is the dimensionless product $T R_3$. We may interpret the limit $T R_3 \rightarrow \infty$ either as a high-temperature limit in fixed volume or as a large-volume limit at fixed temperature. We will be working in a fixed volume, so we will interpret this as a high-temperature limit. In this limit our analysis should agree with any known physics of the finite-temperature SYM theory in flat space. From the supergravity perspective, our analysis should agree with any known physics in Poincar\'{e}-patch $AdS_5$-Schwarzschild.

In the course of our analysis we will study the SYM theory formulated on other curved manifolds besides $S^1 \times S^3$. To do so, we will exploit the fact that global $AdS$ admits various ``slicings.'' To understand these, recall that at the boundary of $AdS$ space the metric has a second-order pole. To extract a boundary metric from a bulk $AdS$ metric, we must choose a defining function which is arbitrary save for one feature: it has a second-order zero at the boundary. Multiplying the bulk metric by this defining function will produce a finite boundary metric. Different slicings of $AdS$ are simply coordinate reparameterizations that naturally suggest different defining functions. Implicitly, we will be using these ``natural'' defining functions, which give rise to boundaries with different geometries. We will be interested in slicings producing the boundary spacetimes%
\footnote
    {%
    The choice of defining function can also determine the topology of the
    boundary.
    For example, a defining function with a zero at some point
    on the boundary may encode a collapsing cycle
    in the boundary spacetime.} $AdS_4$, $AdS_3 \times S^1$, $AdS_2 \times S^2$, $S^1 \times S^3$, and $S^4$. We may write the $AdS_5$ metric for these various slicings as
\begin{equation}
\label{slicingmetric}
ds^2 = d\r^2 + \cosh^{2}\r \; ds^{2}_{AdS_{4-l}} + \sinh^2\r \; d\Omega^{2}_{l}
\end{equation}
with $d\Omega^2_l$ the metric of the unit $l$-sphere $S^l$ and $l = 0, \ldots 4$. When $l \leq 2$, the slicing includes $ds^{2}_{AdS_{4-l}}$, which denotes the $AdS_{4-l}$ metric. We adopt a form for the $AdS_{4-l}$ metric that is identical to Eq.~(\ref{thermaladsmetric}), except for the replacement of the $S^3$ metric, $d\Omega_3^2$, by an $S^{2-l}$ metric, $d\Omega_{2-l}^2$. When $l=3$, we adopt a convention that $AdS_1 \equiv S^1$. Notice that the $S^l$ collapses to zero volume at the center of $AdS$.

Not all of these boundary spacetimes have finite volume, but all introduce a spatial curvature scale into the dual SYM theory, allowing the otherwise scale-free $\Nfour$ SYM theory to undergo a thermal transition. We denote the Ricci scalar of the boundary spacetime as $\mathcal{R}$. For example, in $S^1 \times S^3$ slicing, $\mathcal{R} = 6 / R_3^2$.

The supergravity theory is formulated on the ten-dimensional spacetime $AdS_5 \times S^5$ (or $AdS_5$-Schwarzschild times $S^5$). Convenient forms of the $S^5$ metric that we will use are
\begin{equation}
d\Omega^{2}_{5} = d\th^2 + \sin^{2}\th \; d\Omega^{2}_{{4-j}}
+ \cos^{2}\th \; d\Omega^{2}_{j} \,,
\label{s5}
\end{equation}
where $j$ is any integer from 1 to 4, and $\th$ runs from $0$ to $\pi/2$.

\subsection{Probe \Dp-branes}  \label{probe}

We now introduce $\Nf$ flavor fields in the SYM theory, in the $\Nf \ll \Nc$ limit, which, in the supergravity description, corresponds to introducing \Dp-branes probing the above geometries. Let us momentarily consider a background that is Poincar\'{e}-patch $AdS_5$ times $S^5$, corresponding to the $\N=4$ SYM theory in flat space and at zero temperature. This geometry arises as the near-horizon limit of a number $\Nc \to \infty$ D3-branes \cite{Maldacena:1997re}.

Known supersymmetric embeddings of probe \Dp-branes in this background include D7-branes extended along $AdS_5 \times S^3$ \cite{Karch:2002sh}, D5-branes extended along $AdS_4 \times S^2$ \cite{Karch:2000gx,DeWolfe:2001pq,Erdmenger:2002ex}, and D3-branes extended along $AdS_3 \times S^1$ \cite{Constable:2002xt}. Introducing D7-branes corresponds in the SYM theory to introducing $\N=2$ supersymmetric hypermultiplets that transform in the fundamental representation of the gauge group and that propagate in 3+1 dimensions. Introducing D5-branes corresponds to introducing flavor fields that again transform in the fundamental representation of the gauge group, but that are confined to propagate along a (2+1)-dimensionsal defect. Similarly, introducing D3-branes corresponds to introducing flavor fields confined to propagate along a (1+1)-dimensional defect. More generally, any probe \Dp-brane that is not extended along all of the $AdS_5$  directions will be dual to flavor fields confined to propagate along some defect. In each case, the flavor fields may be given a supersymmetry-preserving mass $m$. In the supergravity description, this mass $m$ is encoded in the geometry of the \Dp-brane in a way that we will make explicit below.

In what follows, most of our attention will be focused on \Dp-branes that are supersymmetric in Poincar\'{e}-patch $AdS_5$ times $S^5$. In particular, we will present numerical results for D7-branes and D5-branes. We would like our analysis to be as general as possible, however, so we will not restrict the $p$ value of the probe \Dp-branes we study.

To introduce the \Dp-brane's induced metric and action, we begin in the low-temperature phase, where the background geometry is global thermal $AdS_5 \times S^5$ (Eqs.~(\ref{thermaladsmetric}) and (\ref{s5})). The \Dp-branes that we will study will be extended along $AdS_i \times S^j$, for some $i$ and $j$ that sum to $p+1$. In other words, the \Dp-brane will be extended along an $AdS_i$ subspace inside global $AdS_5$. This $AdS_i$ subspace includes the radial and time directions as well as an $S^{i-2}$ inside the $S^3$ of $AdS_5$. The \Dp-brane will also be extended along an $S^j$ inside the $S^5$ factor. In the low-temperature phase, we will use the other slicings of global thermal $AdS_5$, Eq.~(\ref{slicingmetric}), only when studying D7-branes, which are extended in all of the $AdS_5$ directions ($i=5$ and $j=3$).

The \Dp-brane may move in directions orthogonal to its worldvolume. We will allow the \Dp-brane to move in the $\theta$ direction defined in Eq.~(\ref{s5}) and, to preserve translation invariance in the dual field theory, we will allow $\theta$ to vary only as a function of the radial coordinate $\rho$. In other words, we allow the position of the $S^j \subset S^5$ to vary as the \Dp-brane extends in the radial direction. From the \Dp-brane worldvolume perspective, $\theta(\rho)$ is a scalar field. As we will review in detail below, this worldvolume scalar is dual to a SYM theory operator that is the (supersymmetric completion of) the mass operator of the flavor fields. This is how the mass $m$ is encoded in the geometry of the \Dp-brane. The induced \Dp-brane worldvolume metric is
\beq
\label{dpmetric}
ds^2_{Dp} = \left[1+\theta'(\rho)^2\right] d\rho^2
+ \cosh^2\!\rho \; d\tau^2 + \sinh^2\!\rho \; d\Omega^2_{{i-2}}
+ \cos^2\!\theta(\rho) \; d\Omega^2_{j}
\eeq
where the prime denotes differentiation with respect to $\rho$. In the high-temperature phase, the embeddings have the same form: $AdS_i$-Schwarzschild times $S^j \subset S^5$, and now the worldvolume scalar $\theta(z)$ depends only on $z$.

To determine $\theta(\rho)$, we need an equation of motion and boundary conditions.  The equation of motion comes from the probe \Dp-brane action, the Dirac-Born-Infeld (DBI) action,\footnote{The full \Dp-brane action of course describes the dynamics of a $U(1)$ worldvolume gauge field and all the worldvolume scalars (including $\theta(\rho)$, but also any others), and includes Wess-Zumino terms that describe the coupling of the \Dp-brane fields to background fields, namely Ramond-Ramond form fields and the Neveu-Schwarz B-field. The equations of motion allow us to set to zero all components of the worldvolume field strength, as well as any other worldvolume scalars (besides our $\theta(\rho)$), in which case we may safely ignore the Wess-Zumino couplings and work with the action in Eq.~(\ref{originaldbi}).}
\beq
\label{originaldbi}
S_{Dp} = N_f \, T_{Dp} \int d^{(p+1)} \zeta \; \sqrt{\det(g_{ab})} \,,
\eeq
where $T_{Dp} = (2\pi)^{-p} g_s^{-1} \alpha'^{-(p+1)/2}$ is the \Dp-brane tension, $\zeta^a$ are worldvolume coordinates and $g_{ab}$ is the induced metric on the \Dp-brane, Eq.~(\ref{dpmetric}). The \Dp-brane metric depends only on $\rho$ (or $z$) and hence $\sqrt{\det(g_{ab})}$ will depend only on $\rho$ (or $z$), so we will rescale the \Dp-brane action and work with an action density. Among the coordinates $\zeta^a$, one will be the radial direction. Another will be the time direction. Integration over this direction simply produces a factor of $1/T$. The \Dp-brane additionally has the internal directions of the $S^j \subset S^5$. Integration over these directions will produce the volume of a unit-radius $j$-sphere, $V_j$. For example, the D7-brane has $V_3 = 2 \pi^2$. The remaining directions are spatial directions inside $AdS_i$. In $S^1 \times S^3$ slicing, these are the directions of the $S^{i-2} \subset S^3$. Integration over these directions then produces a factor of $V_{i-2}$. In the various other slicings, these directions are not all compact, so integration over these directions will not always yield a finite volume. We will divide both sides of Eq.~(\ref{originaldbi}) by the factor of $1/T$ and the integral over the spatial directions inside $AdS_i$, and define an action density
\begin{equation}
\label{dpdbi0}
\tilde{S}_{Dp} \equiv \N_{Dp}
\int d\rho \; \sqrt{\det(g_{ab})} \,.
\end{equation}
where $\N_{Dp} \equiv \Nf T_{Dp} V_j$. From now on, we will refer to $\tilde{S}_{Dp}$ as the \Dp-brane action.

To explain the boundary conditions on the worldvolume scalar, we begin in the high-temperature phase, where the background geometry is global $AdS_5$-Schwarzschild times $S^5$. We begin here because the story is essentially the same as for \Dp-branes in Poincar\'{e}-patch $AdS_5$-Schwarzschild times $S^5$. In the high-temperature phase, two classes of \Dp-brane embedding are possible. In the first class, which we will call ``Minkowski'' embeddings, the $S^j \subset S^5$ that the \Dp-brane wraps is, at the boundary of $AdS_5$-Schwarzschild, the maximum-volume equatorial $S^j$. As the \Dp-brane extends into $AdS_5$-Schwarzschild, the volume of the $S^j$ shrinks (as described by $\theta(z)$) and, indeed, may shrink to zero volume at some radius outside the horizon, $z = \bar z < z_H$. The \Dp-brane does not extend past $\bar z$, rather, it appears to end at $\bar z$ \cite{Karch:2002sh} and does not intersect the $AdS_5$-Schwarzschild horizon. Explicitly, the boundary conditions on $\theta(z)$ are that $\theta(\bar z)=\frac{\pi}{2}$, and $\theta'(\bar z) = \infty$ in order to
avoid a conical singularity \cite{Karch:2006bv}. In the second class of embeddings, which we will call ``black hole'' embeddings, the $S^j \subset S^5$ shrinks as one moves into the bulk of $AdS_5$-Schwarzschild but never collapses to zero volume, so the \Dp-brane intersects the horizon, thus developing a worldvolume horizon. Explicitly, we have $\theta(z_H) \in [0,\frac{\pi}{2})$, and $\theta'(z_H) = 0$ in order for the embedding to be static. These two classes of embedding are distinguished by topology: in Minkowski embeddings the $S^j$ collapses to zero volume, while in black hole embeddings only the thermal circle collapses to zero volume, at the horizon. The ``critical solution'' has $\bar z = z_H$, that is, the critical \Dp-brane ends precisely at the horizon, and \textit{both} the $S^j$ and the thermal circle collapse to zero volume.

As first explained in Ref.~\cite{Karch:2006bv}, analogous embeddings exist in the low-temperature phase, but now the center of thermal $AdS$ plays the role of the horizon. Minkowski embeddings now become ``branes ending away from the center,'' that is, the $S^j \subset S^5$ collapses to zero volume for some nonzero $\bar\rho$, with boundary conditions $\theta(\bar\rho) = \frac{\pi}{2}$ and $\theta'(\bar\rho) = \infty$. Black hole embeddings become ``branes that reach the center,'' that is, the $S^j \subset S^5$ never collapses to zero volume, and the \Dp-brane extends all the way to $\rho = 0$, with boundary conditions $\theta(0) \in [0,\frac{\pi}{2})$ and $\theta'(0)=0$. These two classes of embeddings are again distinguished by topology: for branes ending away from the center, the $S^j \subset S^5$ collapses to zero volume, while for branes that reach the center, the $S^{i-2} \subset AdS_i$ collapses to zero volume. These embeddings are depicted schematically in figure \ref{topchange}. The critical solution is now a \Dp-brane that ends precisely at the center, so $\theta(0) = \frac{\pi}{2}$ and \textit{both} the $S^j$ and $S^{i-2}$ collapse to zero volume at $\rho=0$.

\begin{FIGURE}[t]
{
\centering
\includegraphics[width=0.8\textwidth]{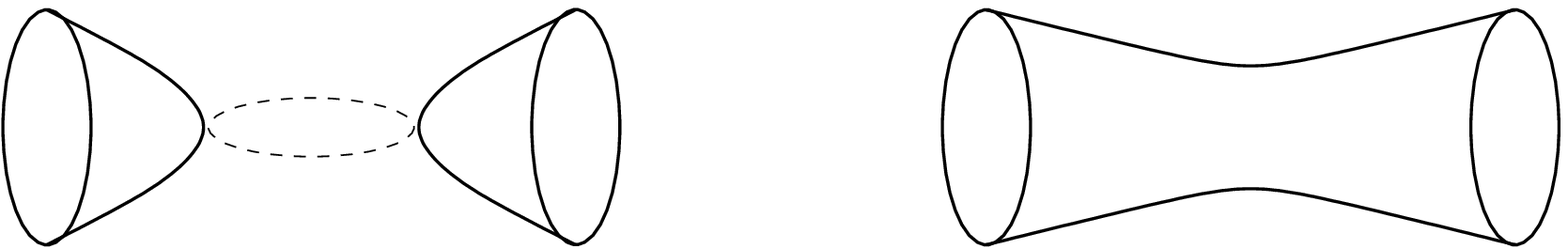}
\caption{\label{topchange} The two topologically distinct \Dp-brane embeddings in thermal $AdS$. The vertical circle represents the $S^j$ factor inside the $S^5$, which can shrink to zero size in the left configuration, the horizontal circle represents the $S^{i-2}$ factor inside of $AdS_i$, which can shrink to zero size in the right configuration.}
}
\end{FIGURE}

We will now explain in detail how the mass $m$ of the flavor fields in the SYM theory, and the thermal expectation value of (the supersymmetric completion of) their mass operator, may be extracted from the \Dp-brane geometry. To be concrete, consider the low-temperature phase. Inserting the worldvolume metric Eq.~(\ref{dpmetric}) into the action Eq.~(\ref{dpdbi0}), we find
\begin{equation}
\tilde{S}_{Dp} = \N_{Dp} \int d\r \> \cosh{\r} \> (\sinh{\r})^{i-2} (\cos{\th})^j \sqrt{1 + \th'^{\,2}}.
\label{dpdbi}
\end{equation}
If we expand this action in a Taylor series in $\th(\rho)$, we may identify, from the $\theta(\rho)^2$ term, a mass-squared for the worldvolume scalar, $M_{\theta}^2 = -j$ (in units of the $AdS$ radius). Probe \Dp-branes fall into two classes depending on whether or not $M_{\theta}^2$ saturates the Breitenlohner-Freedman bound \cite{Breitenlohner:1982jf}, $M^2 \geq -\coeff 14 {(i{-}1)^2}$, for a field in $AdS_i$ with mass-squared $M^2$. The field $\th(\rho)$ for probe D5-branes and D7-branes \textit{does not} saturate the Breitenlohner-Freedman bound, while $\th(\rho)$ for probe D3-branes \textit{does} saturate the Breitenlohner-Freedman bound.

One may relate $M_{\theta}^2$ to the dimension $\D$ of the SYM theory operator dual to $\theta(\rho)$ in the usual fashion, $M_{\theta}^2 = [\D - (i{-}1)]\D$. What is the dual operator? The operator dual to $\theta(\rho)$ is given by the variation with respect to $m$ of the SYM theory Lagrangian. We will denote this operator as $\Om$. For example, consider the probe D7-brane. The dual field theory in this case is $\Nfour$ SYM theory coupled to $\Ntwo$ supersymmetric hypermultiplets in the fundamental representation of the gauge group. The operator $\Om$ is the sum of three terms: the mass operator for the hypermultiplet fermions, $m$ times the mass operator of the hypermultiplet scalars, and a term coupling these scalars to an adjoint scalar. The exact operator is written in Ref.~\cite{Kobayashi:2006sb}. The subscript on $\Om$ is a reminder that this operator depends explicitly on $m$. For our purposes thinking of $\Om$ as the mass operator of flavor fermions will be sufficient (we will not need the explicit form).

The DBI action, Eq.~(\ref{dpdbi}), yields the equation of motion for $\th(\rho)$. Given a solution, we can extract the value of $m$ from the asymptotic behavior of $\theta(\rho)$ \cite{Witten:1998qj,Gubser:1998bc}. The asymptotic behavior is most easily studied (in both the low- and high-temperature phases) using Fefferman-Graham coordinates,
\begin{equation}
\th(z) = z^{i-1-\D} \left\{\th_{(0)} + z \, \th_{(1)} + \cdots
+ z^{2\D - i + 1} \left[ \th_{(2\D - i + 1)} + \ps_{(2\D - i + 1)} \log z\right]
+ \cdots \right\}.
\end{equation}
For \Dp-branes that \textit{do not} saturate the
Breitenlohner-Freedman bound, the leading term, with coefficient
$\th_{(0)}$, is non-normalizable, while the sub-leading term with
coefficient $\th_{(2\D - i +1)}$ is normalizable. The equation of
motion for $\theta(z)$ determines all other coefficients in terms of
these two. These two coefficients will be fixed by the boundary
conditions explained above, which thus completely specify the
asymptotic behavior. In particular, for these cases the coefficient
$\th_{(0)}$ is related to the mass of the fundamental-representation
fields as $m = \th_{(0)}/(2\p\a')$, or equivalently $\theta_{(0)} =
2 \pi \sqrt{\lambda} \, m$. Notice that $\theta_{(0)}$ is
dimensionless and that the right-hand side of this equation is
written in units of the $AdS$ radius. From the SYM theory
perspective, in $S^1 \times S^3$ slicing, the more natural scale is
the radius of the 3-sphere, $R_3$. In our units, $R_3 = 1/2$, so we
may write $\theta_{(0)}= 4 \pi m R_3 / \sqrt{\lambda}$.

For \Dp-branes that \textit{do} saturate the Breitenlohner-Freedman bound, which have $2\D = i{-}1$, the coefficients $\th_{(0)}$ and $\th_{(2\D - i +1)}$ are identical. The leading logarithmic term, with coefficient $\ps_{(2\D - i + 1)} = \psi_{(0)}$, is now non-normalizable, while the term with coefficient $\theta_{(0)}$ is normalizable. In the general analysis of section \ref{scaling}, we will use the notation appropriate for \Dp-branes that \textit{do not} saturate the Breitenlohner-Freedman bound, namely $\theta_{(0)}$ as the coefficient of the non-normalizable term and $\theta_{(2\D - i +1)}$ as the coefficient of the normalizable term. Converting to \Dp-branes that \textit{do} saturate the Breitenloner-Freedman bound is easy: simply replace $\theta_{(0)} \ra \psi_{(0)}$ and $\theta_{(2\D - i + 1)} \ra \theta_{(0)}$.

Following Ref.~\cite{Karch:2005ms}, we can extract the thermal expectation value $\lang \Om \rang$ from the asymptotic coefficients using holographic renormalization. Let $F$ denote the contribution that the flavor fields make to the free energy density. $\lang \Om \rang$ is given by $\frac{\delta F}{\delta m}$. Using the AdS/CFT correspondence, $F$ is given by the value of the on-shell DBI action Eq.~(\ref{dpdbi0}), so we need to take the variation of the on-shell DBI action with respect
to the leading, non-normalizable asymptotic value of $\theta(z)$. The problem that arises in doing this is that the radial integration will diverge at the boundary, $z=0$. In holographic renormalization, we introduce a regulator by cutting off the integration at some small value, \textit{i.e.}, we integrate only to $z = \epsilon$. We then introduce counterterms on the $z = \epsilon$ hypersurface to cancel divergences before removing the regulator by sending $\epsilon \ra 0$. These counterterms are written in terms of $\theta(\epsilon)$ and the induced metric on the $z = \epsilon$ hypersurface, whose determinant we denote as $\g$. We denote the regulated action, plus counterterms, as $\tilde{S}_{reg}$, so that $\tilde{S}_{reg} = F$. We then have, for \Dp-branes that \textit{do not} saturate the Breitenlohner-Freedman bound \cite{Karch:2005ms,deHaro:2000xn}, \begin{equation}\label{dpvev}
\langle {\Om} \rangle = \lim_{\e \rightarrow 0} \left(\frac{\e^{-\D}}
{\sqrt{\g}} \>  \frac{\d \tilde{S}_{\rm reg}}{\d \th(\e)}\right).
\end{equation}
For \Dp-branes that \textit{do} saturate the Breitenlohner-Freedman bound, we must make the replacement $\e^{-\D} \rightarrow \e^{-\D} \log \e$ \cite{Karch:2005ms}. The counterterms and explicit formulae for $\lang \Om \rang$ for the D7-brane, D5-brane and D3-brane are given in Refs.~\cite{Karch:2005ms,Karch:2006bv}.

Let us illustrate the procedure with an example: the probe D7-brane, which is extended along $AdS_5 \times S^3$ and so has $i=5$, $j=3$ and $M_{\theta}^2 = -j = -3$. The D7-brane thus \textit{does not} saturate the Breitenlohner-Freedman bound for $AdS_5$, $M^2 \geq - 4$. The dual operator $\Om$ has dimension $\Delta = 3$. $\th(z)$ has an asymptotic expansion
\begin{equation}
\th(z) = \th_{(0)} \, z + \th_{(2)} \, z^3 + \psi_{(2)} \, z^3 \log z + \cdots
\,,
\end{equation}
where $\psi_{(2)}$ is fixed in terms of $\th_{(0)}$ as $\psi_{(2)}
= \frac{1}{12} {\cal{R}} \, \th_{(0)}$. The regulated action is $\tilde{S}_{reg} = \tilde{S}_{D7} + \sum_k \tilde{L}_k$, where the counterterms are \cite{Karch:2005ms,Karch:2006bv}
\begin{subequations}
\label{d7counterterms}
\begin{align}
\tilde{L}_1 &= - \coeff{1}{4} \sqrt{\g}\,,&
\tilde{L}_2 &= \coeff{1}{48} \sqrt{\g} \, {\cal{R}}_{\g}\,,&
\tilde{L}_3 &= -  \coeff{1}{32} (\log \e)\sqrt{\g}
\left(
{\cal{R}}^{\gamma}_{ij} \,{\cal{R}}_{\gamma}^{ij} - \coeff{1}{3}\, {\cal{R}}_{\g}^2
\right) ,
\\
\tilde{L}_4 &= \coeff{1}{2} \sqrt{\g} \, \th(\e)^2\,,&
\tilde{L}_5 &= -\coeff{5}{12} \sqrt{\g} \, \th(\e)^4 \,,&
\tilde{L}_6 &= \coeff{1}{12} \log\th(\e) \sqrt{\g} \, {\cal{R}}_{\g} \, \th(\e)^2\,.
\end{align}
\end{subequations}
where for clarity we have suppressed a factor of $\N_{D7} = N_f T_{D7} V_3 = \frac{1}{(2\pi)^4} \lambda N_f N_c$ that appears in every counterterm. Here ${\cal{R}}^{\gamma}_{ij}$ and ${\cal{R}}_{\g}$ are the Ricci tensor and  Ricci scalar, respectively, of the induced metric
on the $z = \epsilon$ hypersurface. Using Eq.~(\ref{dpvev}), we find that the thermal expectation value of $\Om$
is given by
\begin{equation}
\label{d7vev}
\langle \Om \rangle = \N_{D7} \left [ - 2 \th_{(2)} + \coeff{1}{3} \th_{(0)}^3 + \coeff{1}{6} {\cal{R}} \, \th_{(0)} \log \th_{(0)} \right ] \,.
\end{equation}

We will also present numerical results for a probe D5-brane, which is extended along $AdS_4 \times S^2$ and so has $i=4$ and $j=2$. The D5-brane has $M_{\theta}^2 = - j = -2$ and hence \textit{does not} saturate the Breitenlohner-Freedman bound for $AdS_4$, $M^2 \geq - \frac{9}{4}$. The dual operator $\Om$ has dimension $\Delta = 2$. In this case, the coefficient of the sub-leading, normalizable term is $\th_{(2\D - i +1)} = \theta_{(1)}$. We will not present the counterterms explicitly. The result for the thermal expectation value of $\Om$ is $\langle \Om \rangle = - \N_{D5} \theta_{(1)} $ \cite{Karch:2005ms}, where $\N_{D5} = \Nf T_{D5} V_2 = \frac{1}{2\pi^3} \sqrt{\lambda} \, \Nf \Nc$.

\section{Scaling Analysis}  \label{scaling}

In this section we perform a scaling analysis similar to that of Refs.~\cite{Frolov:2006tc,Mateos:2006nu,Mateos:2007vn} for probe \Dp-branes in global thermal $AdS$, probe \Dp-branes in global $AdS$-Schwarzschild, and probe D7-branes in the various other slicings of global thermal $AdS$. We find the relevant scaling exponents for probe \Dp-brane embeddings and determine when those exponents are real or complex. We also confirm via numerical analysis that complex exponents
signal a first order transition in the dual SYM theory.

\subsection{Probe \Dp-branes in $S^1 \times S^3$ Slicing: Low-temperature Phase}  \label{low}

We begin with probe \Dp-branes in global thermal $AdS$, corresponding to the SYM theory in the low-temperature, confined phase. We first note that the flavor physics, within the low temperature phase of the SYM theory, is completely independent of temperature. The global thermal $AdS$ metric Eq.~(\ref{thermaladsmetric}) does not depend on the temperature $T$, which thus appears in the DBI action Eq.~(\ref{originaldbi}) only via an overall factor from
integration over the thermal $S^1$ (which is then hidden in the rescaled action Eq.~(\ref{dpdbi0})). Consequently, the equation of motion for $\th(\r)$, and its boundary conditions, are $T$-independent. As a result, the expectation value $\lang \Om \rang$ in the dual SYM theory will be independent of $T$. This temperature independence is not unexpected --- it is a property of the leading large-$\Nc$ behavior of generic observables (not involving Wilson loops which wrap the thermal circle) in the low temperature phase of non-Abelian gauge theories \cite{Neri:1983ic,Pisarski:1983db,Kovtun:2007py}. Our supergravity analysis is thus valid for any temperature below the Hawking-Page transition temperature, $T < T_{HP} = \frac{3}{2\pi}$. The remaining relevant scales in the SYM theory are the hypermultiplet mass $m$ and the radius $R_3$ of the 3-sphere. The physics we are interested in will depend only on these scales, or more accurately on their dimensionless product $m R_3$.

We now turn to the scaling analysis. We focus on the region near the center of $AdS$, and expand the metric to leading nontrivial order using $\sinh \r \approx \r$, $\cosh \r \approx 1$.
We first consider the critical solution, that, is, fluctuations of the form $\th(\r) = \pi/2 + \d\th(\r)$ with $\d\th(\r)$ small. We may then use $\cos{\th} \approx - \d\th$. The critical solution has the boundary condition $\d\th(0)=0$. The \Dp-brane action, Eq.~(\ref{dpdbi}), becomes
\begin{equation}
\tilde{S}_{Dp} = \N_{Dp} \int d\r \> \r^{i-2} \; \d\th^j \sqrt{1+\d\th'^{\,2}} \,,
\label{ncdbi}
\end{equation}
with, once again, $i+j = p+1$.
The resulting equation of motion for $\d\th(\r)$ is
\begin{equation}
\r \, \d\th \, \d\th'' +\left[(i{-}2)\d\th \, \d\th' - j\r\right](1+\d\th'^{\,2}) = 0\,.
\label{deltathetaeom}
\end{equation}
This equation has an important scaling symmetry under which
\begin{equation}
\d\th(\rho) \ra \m \, \d\th(\rho) \quad\hbox{and}\quad
\r \ra \m \r  \,,
\label{eq:scalingtrans}
\end{equation}
for real, positive $\mu$. In other words, a single solution $\d\th(\rho)
= f(\rho)$ gives rise to a one-parameter family of solutions $\d\th(\rho)
= \mu^{-1} {f(\mu \rho)}$. The solution of Eq.~(\ref{deltathetaeom})
for the critical embedding, which we denote $\d \theta^*(\rho)$, is
\begin{equation}
\d\th^*(\r) \equiv \r \> \sqrt{\frac{j}{i{-}2}} \,.
\end{equation}
Notice that $\d \th^*(\rho)$ is invariant under the scaling transformation.

\begin{TABLE}
{
\renewcommand\arraystretch{1.5}
\quad\begin{tabular}{|c|c|}
\hline
$a$ & $\b_{\pm}$ \\
\hline
7 & $-3 \pm \sqrt{2}$ \\
\hline
6 & $-\frac{5}{2} \pm \frac{1}{2}$ \\
\hline
5 & $-2 \pm i$\\
\hline
4 & $-\frac{3}{2} \pm i \frac{\sqrt{7}}{2}$ \\
\hline
3 & $-1 \pm i \sqrt{2}$\\
\hline
2 & $-\frac{1}{2} \pm i \frac{\sqrt{7}}{2}$\\
\hline
\end{tabular}\quad
\caption{\label{exponents} Scaling exponents $\b_{\pm}$ for probe
\Dp-branes whose critical solution has a collapsing submanifold of
dimension $a$.
\vspace*{30pt}
}
}
\end{TABLE}
Next, we seek solutions near the critical one, still in the near-center region, of the form
\begin{equation}
\d\th(\r) = \d\th^*(\r) + \xi(\r) \,.
\end{equation}
The fluctuation $\xi(\r)$ has the linearized equation of motion
\begin{equation}
\r^2 \, \xi'' + a \, \r \, \xi' + a \, \xi = 0 \,,
\end{equation}
with $a = j+ (i -2)$. This has solutions $\xi(\r) = \pm\r^{\b_{\pm}}$ with
\begin{equation}
\b_{\pm} = \half \left [ (1-a) \pm \sqrt{a^2 - 6a + 1} \right ].
\end{equation}
The $\beta_{\pm}$ are the scaling exponents.
Depending on the value of $a$,
the scaling exponents $\b_{\pm}$ may be real or complex.
Physically, $a = j + (i-2)$
is the dimension of the sub-manifold of the \Dp-brane worldvolume that
collapses to zero volume in the critical solution: the $j$ dimensions of
the $S^j \subset S^5$ and the $i{-}2$ dimensions of the $S^{i-2} \subset
AdS_i$.
For the present case of probe D$p$-branes in $S^1 \times S^3$ slicing,
the dimension $a$ of the collapsing submanifold is simply $p{-}1$,
but this will not be the case in subsequent examples.

Explicit values of the exponents, for various choices of $a$,
appear in Table \ref{exponents}.
As shown in the table, the critical value of $a$ is 6:
for $a<6$ the near-center, near-critical solutions have complex
exponents, while for $a \geq 6$ they have real exponents.

The perturbed solution,
\begin{equation}
\d\th(\r) = \d\th^*(\r) + \a_+ \> \r^{\b_+} + \a_- \> \r^{\b_-} \,,
\end{equation}
must obey the scaling symmetry. This implies that the coefficients $\a_{\pm}$ must scale as
\begin{equation}
\label{alphascaling}
\a_{\pm} \ra \m^{1-\b_{\pm}} \; \a_{\pm} \,.
\end{equation}
The coefficients $\a_{\pm}$ are determined by the boundary conditions, specifically, the value of $\bar{\rho}$ or $\theta(0)$. This is why $\a_{\pm}$ are ``charged'' under the scaling transformation: because $\bar{\rho}$ and $\theta(0)$ are  not invariant under scaling.

A solution that is near the critical solution in the near-center region will remain so all the way out to the asymptotic region.\footnote{More formally, we expect the critical solution to be an attractor solution as one moves toward the $AdS$ boundary. In Ref.~\cite{Frolov:2006tc}, this was shown explicitly for \Dp-branes in Poincar\'{e}-patch $AdS$-Schwarzschild.} The boundary condition deep inside $AdS_i$ (the value of $\bar{\rho}$ or $\th(0)$) must fix both the asymptotic coefficients $\th_{(0)}$ and $\th_{(2\Delta -i + 1)}$ and the near-center coefficients $\a_{\pm}$. The asymptotic coefficients can thus be thought of as functions of the near-center coefficients. Sufficiently close to the critical solution, the $\a_{\pm}$ will be very small and hence the asymptotic coefficients may be linearly related to the near-center coefficients. Put more simply, we may Taylor expand $\theta_{(0)}$ and $\theta_{(2\Delta -i + 1)}$ to linear order in $\alpha_{\pm}$. Let $\theta_{(0)}^*$ and $\theta_{(2\Delta -i + 1)}^*$ denote the asymptotic coefficients of the critical solution. We then have, for a near-critical solution,
\begin{subequations}
\label{coeffdiffs}
\begin{align}
\th_{(0)} - \th_{(0)}^* &= A_{(0)}^{+} \; \a_{+} + A_{(0)}^{-} \; \a_- \,,
\\
\th_{(2\Delta -i + 1)} - \th_{(2\Delta -i + 1)}^*
&= A_{(2\Delta -i + 1)}^{+} \; \a_{+} + A_{(2\Delta -i + 1)}^{-} \; \a_- \,,
\end{align}
\end{subequations}
for some set of coefficients $A^{\pm}$. Notice that the overall sign of $\xi(\rho)$ is arbitrary, so the form of Eq. (\ref{coeffdiffs}) is valid for both $\theta_{(0)} < \theta_{(0)}^*$ and $\theta_{(0)} > \theta_{(0)}^*$, that is, for \Dp-branes that end at the center and for \Dp-branes that end away from the center. We may now ask what the scaling transformation of the $\alpha_{\pm}$, Eq.~(\ref{alphascaling}), teaches us about the asymptotic coefficients, and what we may then conclude in the SYM theory about the behavior of $\lang \Om \rang$ as a function of $m$.

For complex $\b_{\pm}$, if we perform the scaling transformation
Eq.~(\ref{alphascaling}) we find that, as functions of $\m$, $(\th_{(0)}
- \th_{(0)}^*)$ and $(\th_{(2\Delta -i + 1)} - \th_{(2\Delta -i +
1)}^*)$ will acquire terms of the form
$\m^{1-\Re \b_{\pm}}$ times either a sine or cosine of
$\Im \b_{\pm} \log \m$. Notice that $1-\Re \b_\pm > 0$ in all cases.
Consequently, sending $\mu \to 0$ produces solutions which,
in the $(\th_{(0)},\th_{(2\Delta -i + 1)})$ plane,
spiral inward toward a limit point which corresponds to
the critical solution. More precisely, the arbitrary overall sign of $\xi(\rho)$
leads to two intertwined spirals, as illustrated below
in Fig.~\ref{d5coeffs}(b).

In the dual field theory, we may argue, following Refs.~\cite{Mateos:2006nu,Mateos:2007vn}, that $\lang \Om \rang$ must jump discontinuously from the outer branch of one spiral to the outer branch of the other as the mass $m$ is varied.
Such discontinuous behavior signals a first-order phase transition. We then expect discontinuous behavior in generic observables associated with the flavor fields, of which $\lang \Om \rang$ is simply the most convenient to compute from the supergravity perspective.

To be concrete, let us analyze the example of probe D5-branes extended along $AdS_4 \times S^2$, for which $a=4$ and $\b_{\pm} = - \frac{3}{2} \pm i \frac{\sqrt{7}}{2}$. Recall that in this case $m = \theta_{(0)}/(2\pi \alpha')$, and $\lang \Om \rang = - \N_{D5} \theta_{(1)}$ where the operator $\Om$ has dimension $\Delta = 2$ and $\N_{D5} = \frac{1}{2\pi^3} \lambda^{1/2} \Nf \Nc$. Figure~\ref{d5action} shows the value of $\tilde{S}_{reg}/\N_{D5}$ for the $D5$-brane as a function of $\th_{(0)}$, and figure~\ref{d5coeffs} shows the numerical result for solutions in the $(\th_{(0)},\th_{(1)})$ plane.

\begin{FIGURE}
{
\includegraphics[width=0.48\textwidth]{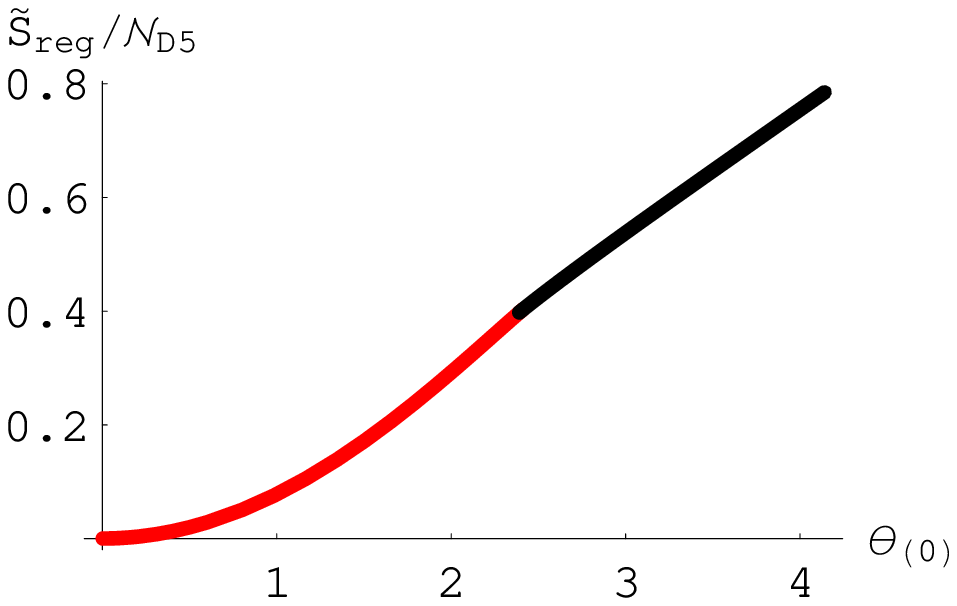} \hfil
\begin{overpic}[width=0.48\textwidth] {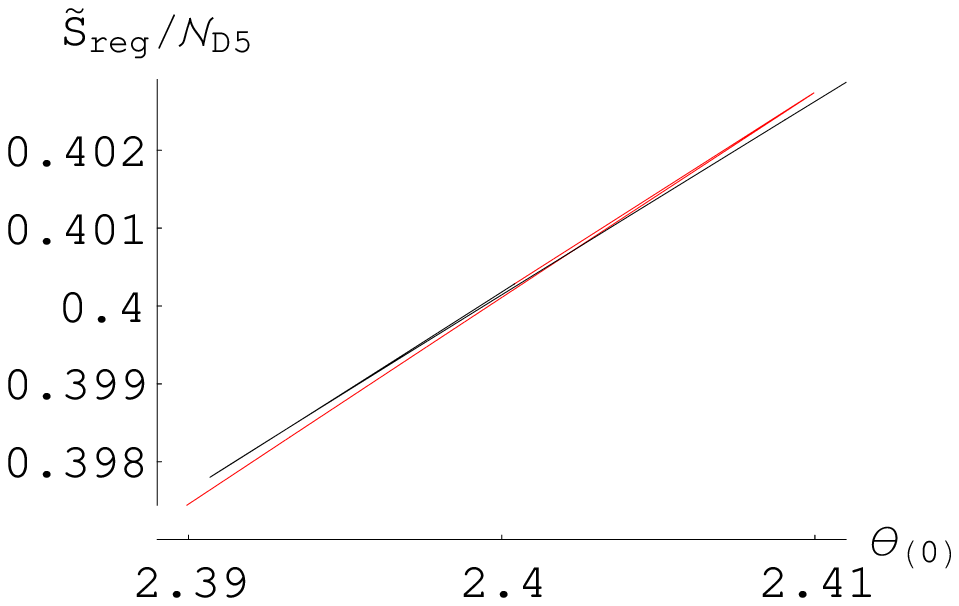} \put(52,6.8){\includegraphics[width=0.24\textwidth]{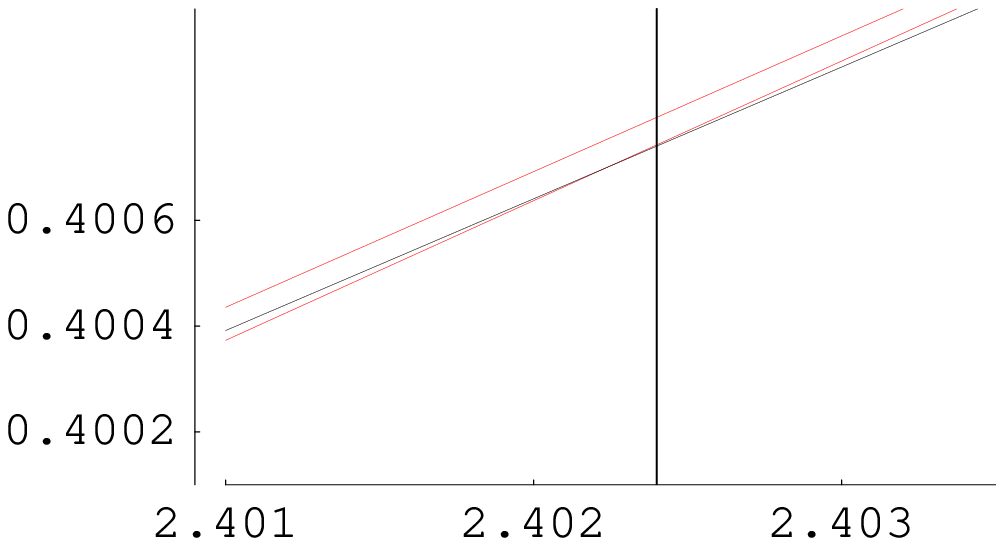}} \end{overpic}
(a) \hfil\hfil (b)
\caption{\label{d5action}(a.) $\tilde{S}_{reg}/\N_{D5}$ as a function of $\th_{(0)}$ for the D5-brane probe in global thermal $AdS$ in $S^1 \times S^3$ slicing. The red curves correspond to D5-branes that reach the center of $AdS$ while the black curves correspond to D5-branes that end away from the center. The point at which the red and black curves meet corresponds to the critical solution. (b.) Close-up of (a.) near the critical solution. The inset figure shows a further close-up. The vertical line in the inset figure indicates where the transition occurs, at the critical value $\theta_{(0)}^{crit} \approx 2.402$.}
}
\end{FIGURE}

\begin{FIGURE}
{
\includegraphics[width=0.48\textwidth]{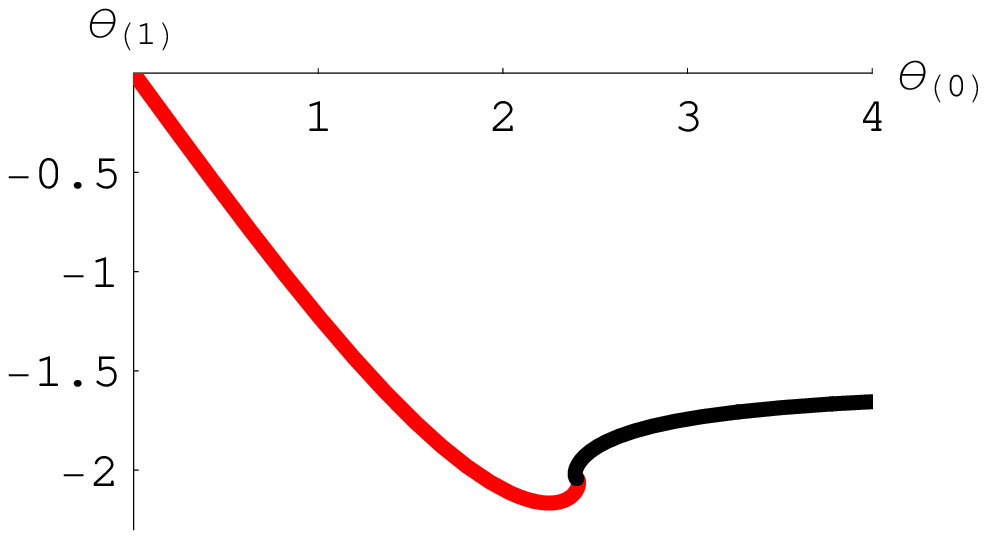} \hfil
\includegraphics[width=0.48\textwidth]{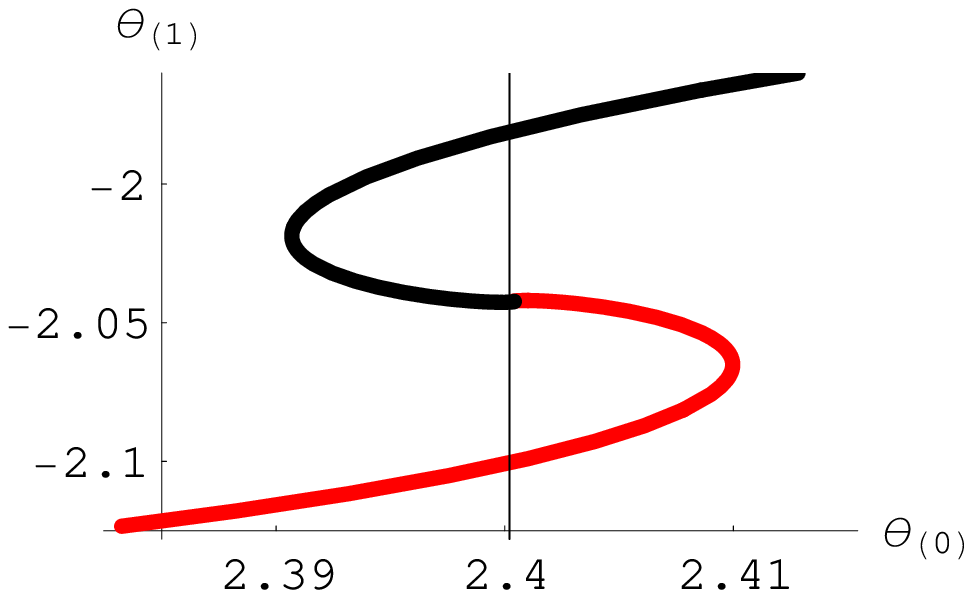}\\
(a) \hfil\hfil (b)
\caption{\label{d5coeffs}(a.) $\th_{(1)}$ as a function of $\th_{(0)}$ for the D5-brane probe in global thermal $AdS$ in $S^1 \times S^3$ slicing. The red curves correspond to D5-branes that reach the center of $AdS$ while the black curves correspond to D5-branes that end away from the center. The point at which the red and black curves meet corresponds to the critical solution. (b.) Close-up of (a.) near the critical solution. The vertical line indicates where the transition occurs, at the critical value $\theta_{(0)}^{crit} \approx 2.400$. As we increase $\theta_{(0)}$, moving along the red curve from left to the right, the physical value of $ \theta_{(1)} = - \langle \Om \rangle / \N_{D5}$ jumps upward at $\theta_{(0)}^{crit}$, from the red curve to the top-most arm of the black curve.
}
}
\end{FIGURE}

In these and all subsequent plots for global thermal $AdS$, the red curves arise from \Dp-brane solutions that reach the center of $AdS$, while the black curves arise from \Dp-branes ending away from the center. For global $AdS_5$-Schwarzschild, red curves arise from \Dp-branes that reach the horizon while black curves arise from \Dp-branes that end outside the horizon. The critical solution will thus always be where the red and black curves meet.

In figure \ref{d5action}, we see that the free energy, which is always the minimal value of the regulated action $\tilde{S}_{reg}$, has a small kink (a discontinuous first derivative) as a function of $\theta_{(0)}$. Since $m = \theta_{(0)}/(2\p\a')$, we conclude that a first-order phase transition occurs in the SYM theory as a function of the mass $m$. (The free energy is always continuous, but at a first order transition
its first derivative jumps.) The transition occurs at the critical value\footnote{Notice that the result for $\theta_{(0)}^{crit}$ as computed from the free energy in figure  \ref{d5action} is $\theta_{(0)}^{crit} \approx 2.402$, which is slightly larger than the value $\theta_{(0)}^{crit} \approx 2.400$ obtained from figure \ref{d5coeffs}, where we used an equal-area method. We attribute the difference to numerical error in estimating the location of the ``kink'' in figure \ref{d5action}. Here, and for all subsequent branes, the location of the transition that we will use is that computed from the equal-area method.} $\theta_{(0)}^{crit} = 4 \pi m R_3/\sqrt\lambda = 2.400$ or equivalently $m = \sqrt\lambda\, \theta_{(0)}^{crit}/(4\pi R_3) = 0.191 \, \sqrt\lambda / R_3$. Using $\lang \Om \rang = \frac{\delta F}{\delta m} = -\N_{D5} \theta_{(1)}$, we see precisely this behavior in figure \ref{d5coeffs}, as $\theta_{(1)}$ jumps discontinuously from one arm of the spiral to the other. When the embedding equations have multiple solutions for a given value of $\theta_{(0)}$, only those solutions that minimize the free energy ($\tilde{S}_{reg}$) represent genuine equilibrium states. These minimal free-energy solutions only lie on the outermost branches of the two spiral arms. As shown in Refs.~\cite{Mateos:2007vn,Filev:2007qu}, moving inward along the red or black curves toward the critical solution, a new tachyon appears in the meson spectrum at every turn of the spiral, providing a clear signal of instability. Notice also that $m^* \sim 1/R_3$, as expected from the fact that generic observables will be $T$-independent in the low-temperature phase, as mentioned above. Clearly this transition is a finite-volume effect.

\begin{FIGURE}[t]
{
\includegraphics[width=0.48\textwidth]{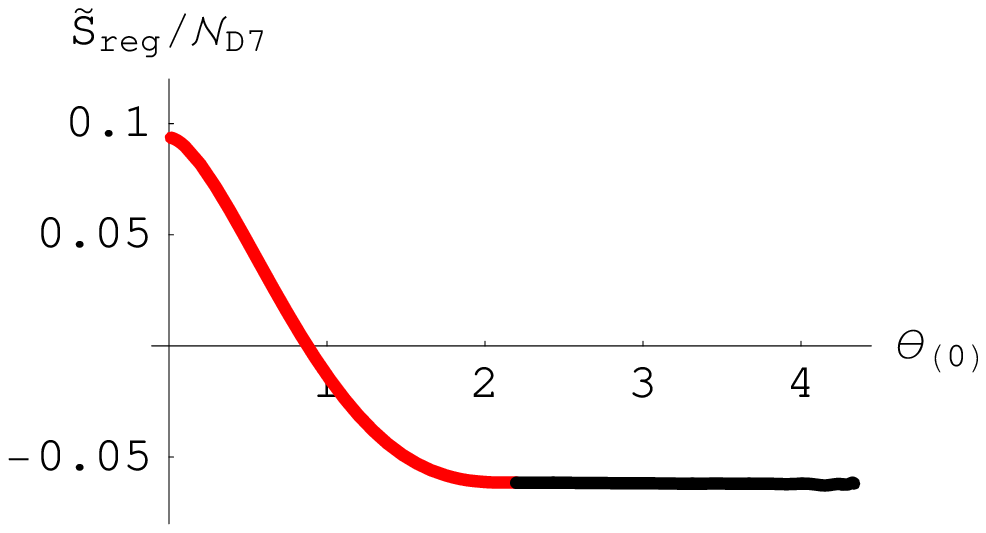} \hfil
\includegraphics[width=0.48\textwidth]{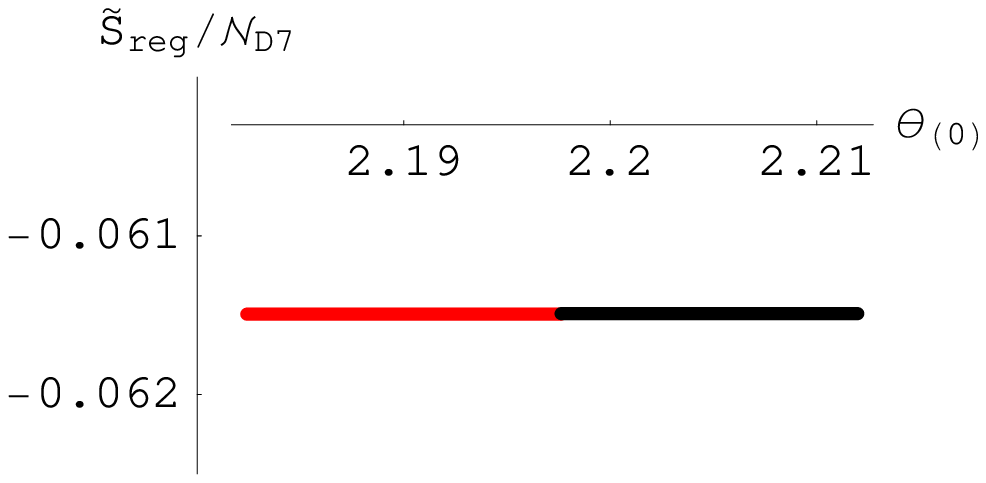}\\
(a) \hfil\hfil (b)
\caption{\label{d7action}(a.) $\tilde{S}_{reg}/\N_{D7}$ as a function of $\th_{(0)}$ for the D7-brane probe in global thermal $AdS$ in $S^1 \times S^3$ slicing. (b.) Close-up of (a.) near the critical solution. The critical solution (where the red and black curves meet) has $\theta_{(0)}^* \approx 2.198$.}
}
\end{FIGURE}

\begin{FIGURE}[t]
{
\includegraphics[width=0.48\textwidth]{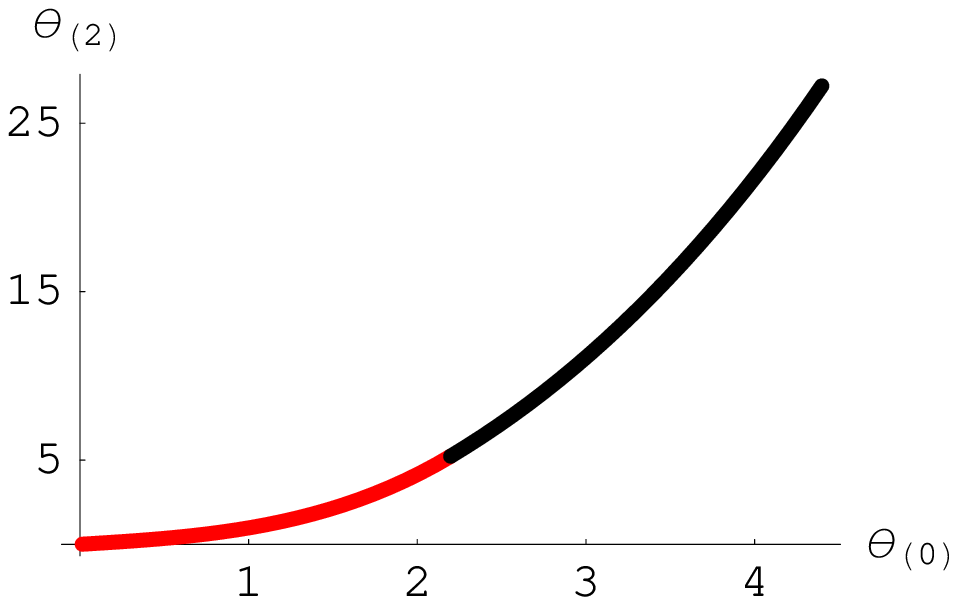} \hfil
\includegraphics[width=0.48\textwidth]{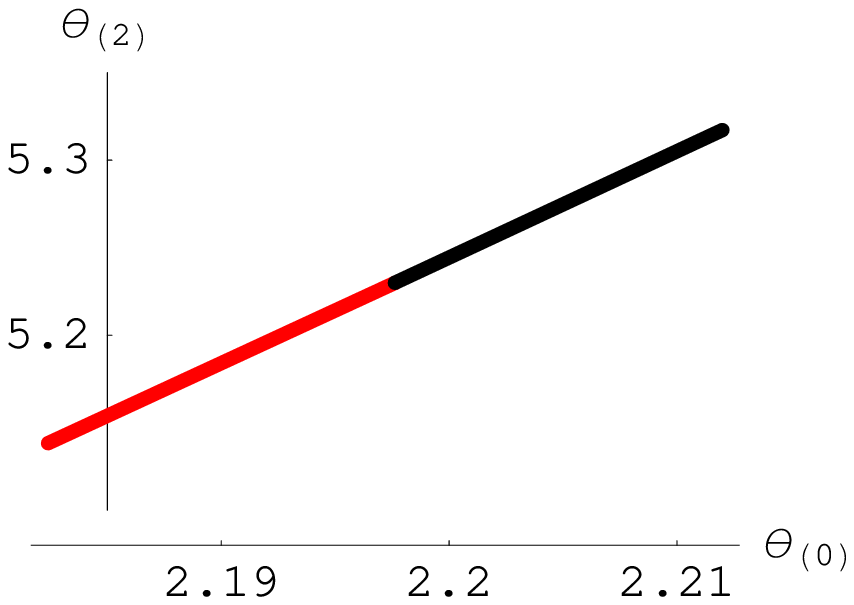}\\
(a) \hfil\hfil (b)
\caption{\label{d7coeffs}(a.) $\th_{(2)}$ as a function of $\th_{(0)}$ for the D7-brane probe in global thermal $AdS$ in $S^1 \times S^3$ slicing. (b.) Close-up of (a.) near the critical solution. The critical solution (where the red and black curves meet) has $\theta_{(0)}^* \approx 2.198$ and $\theta_{(2)}^* \approx 5.230$.}
}
\end{FIGURE}

When the scaling exponents $\beta_{\pm}$ are real, the story is very different. Now the scaling transformation of Eq.~(\ref{alphascaling}) is a simple rescaling, and no spiral will appear in the $(\th_{(0)},\th_{(2\Delta -i + 1)})$ plane. The most interesting example is the probe D7-brane,%
\footnote{We have also formally analyzed a D8-brane extended along $AdS_5 \times S^4$, which has $a=7$, and found behavior similar to that for the D7-brane. A D8-brane carries no conserved charge in type IIB supergravity, however, and is therefore unstable.
}
extended along $AdS_5 \times S^3$, for which $a=6$ and $\b_{\pm}= - \frac{5}{2} \pm \frac{1}{2}=\{-2,-3\}$. Recall that in this case $m = \theta_{(0)}/(2\pi \alpha')$ and $\langle \Om \rangle$ is given by Eq.~(\ref{d7vev}) where the operator $\Om$ has dimension $\Delta = 3$. Recall also that $\N_{D7} = \frac{1}{(2\pi)^4} \lambda N_f N_c$. Figure \ref{d7action} shows $\tilde{S}_{reg}/\N_{D7}$ as a function of $\theta_{(0)}$ for the D7-brane, which has a continuous first derivative. Figure \ref{d7coeffs} shows $\theta_{(2)}$ as a function of $\theta_{(0)}$, which is single-valued. The critical solution has $\theta_{(0)}^* \approx 2.198$ and $\theta_{(2)}^* \approx 5.230$. In terms of SYM theory quantities, the critical solution has $m^* = \sqrt{\lambda} \, \theta_{(0)}^* / (4\pi R_3) = 0.1748 \sqrt{\lambda} / R_3$. (Notice again that $m^* \sim 1/R_3$.) We conclude that no first-order transition occurs in the SYM theory. We expect some kind of non-analyticity in the SYM theory, however, because in the supergravity description the topology of the D7-brane changes in passing through the critical solution. In section \ref{transition} we will show that certain observables have divergent behavior at the critical point, and we will examine how this transition affects the meson spectrum.

\subsection{Probe \Dp-branes in $S^1 \times S^3$ Slicing: High-temperature Phase}  \label{high}

We will now perform the scaling analysis for probe \Dp-branes in global $AdS_5$-Schwarzschild, corresponding to the SYM theory in the high-temperature, deconfined phase. Our result will be essentially the same as that of Refs.~\cite{Mateos:2006nu,Mateos:2007vn}: all probe \Dp-branes have complex exponents $\b_{\pm}$ and hence exhibit a spiral in the $(\th_{(0)},\th_{(2\Delta -i + 1)})$ plane, implying that first-order transitions appear in the respective SYM theories. For the D7-brane, this must be the case in order to agree with the numerical results in $S^1 \times S^3$ slicing
\cite{Karch:2006bv} and in Poincar\'{e} patch $AdS_5$-Schwarzschild
\cite{Babington:2003vm,Kirsch:2004km,Ghoroku:2005tf,Albash:2006ew,Mateos:2006nu,Karch:2006bv,Mateos:2007vn},
which we interpret as the high-$T$ limit of the theory formulated on global $AdS_5$-Schwarzschild, as explained in section \ref{background}.

The critical solution is now a \Dp-brane that ends precisely at the horizon. At the horizon only the thermal $S^1$ inside
$AdS_i$-Schwarzschild collapses to zero volume. This is the key difference from thermal $AdS_i$. For the critical solution, the $S^j \subset S^5$ will still collapse, but now \textit{all} probe \Dp-brane critical embeddings will have the same collapsing $S^1$. The $S^{i-2}$ that collapsed in thermal $AdS_i$ are replaced with this $S^1$ in $AdS_i$-Schwarzschild, so we expect $a = j+1$. We can thus jump to the answer: as the largest value of $j$ that allows for $\theta(z)$ is $j=4$, none of the \Dp-brane probes can have $a > 6$, and hence all must have complex $\beta_{\pm}$.

To confirm this, we proceed in the same spirit as above. We now focus
on the near-horizon region. Let $z = z_H + Z$
and $\theta(z) = \frac \pi 2 + \delta\theta(Z)$.
We expand the metric
coefficients to leading nontrivial order. The induced D$p$-brane metric
is then
\begin{equation}
ds^{2}_{Dp} = \left ( 1 +  \delta \theta'^{\,2} \right ) dZ^2 + \frac{2}{z_{H}^2} \frac{Z^2 }{1-z_{H}^2} \; d\t^2 +  {\mathcal{F}}(z_H) \; d\Omega^2_{{i-2}} + \delta \theta^2 \, d\Omega^2_{j}.
\end{equation}
The \Dp-brane action, ignoring overall $Z$-independent constants that do not affect $\delta \theta$'s equation of motion, is
\begin{equation}
\label{adsbhdbi}
S_{Dp} \propto \int dZ Z \> \delta \theta^j \sqrt{1 + \delta \theta'^{\,2}} \,.
\end{equation}
This is of precisely the same form as Eq.~(\ref{ncdbi}) but with
$i-2 \ra 1$ so indeed $a = j+1$ and all D$p$-brane probes will have
complex exponents $\beta_{\pm}$ and exhibit a spiral in the
$(\th_{(0)},\th_{(2\Delta -i + 1)})$ plane. The action
Eq.~(\ref{adsbhdbi}) is in fact identical to the near-horizon action
written in Refs.~\cite{Mateos:2006nu,Mateos:2007vn} for D$p$-brane
probes in the Rindler space that arises as the near-horizon geometry
of Poincar\'{e} patch $AdS$-Schwarzschild.

\subsection{Probe D7-brane in Other Slicings: Low-Temperature Phase}  \label{other}

By using different slicings, leading to different boundary geometries, we can change the dimension of the critical solution's collapsing submanifold, as different slicings lead to spheres of different dimension, $S^l \subset AdS_5$ for $l = 0,\ldots,4$, that collapse to zero volume at the center of $AdS_5$ (see Eq.~(\ref{slicingmetric})). We will focus on the D7-brane and the low-temperature phase because lower-dimensional \Dp-branes, or \Dp-branes in the high-temperature phase, will not be able to reach $a \geq 6$. For the D7-brane, the dual SYM theory is $\Nfour$ SYM theory coupled to massive $\Ntwo$ hypermultiplets, formulated on different four-manifolds, in the low-temperature, confining phase.

Using the $AdS_{4-l} \times S^l$ slicing of $AdS_5$,
the induced D7-brane metric is
\beq
ds^2_{D7} = \left[1+\theta'(\rho)^2\right] d\rho^2
+ \cosh^2\!\rho \> ds^2_{AdS_{4-l}}
+ \sinh^2\!\rho \> d\Omega^2_{l}
+ \cos^2\theta(\rho) \> d\Omega^2_{3} \,,
\eeq
and the D7-brane action is
\begin{equation}
\tilde{S}_{D7} = \N_{D7} \int d\r \> (\cosh \r)^{4-l} \, (\sinh \r)^l \, (\cos\th)^3 \sqrt{1 + \th'^{\,2}} \,.
\end{equation}
In the near-center limit this becomes
\begin{equation}
\tilde{S}_{D7} = \N_{D7} \int d\r \> \r^l \, \delta \theta^3 \sqrt{1+\delta \theta'^{\,2}}
\end{equation}
which is the same as Eq.~(\ref{ncdbi}), but with $i - 2 \ra l$ and $j = 3$, so $a = l +3$ and we will have complex exponents $\beta_{\pm}$ for $l = 0,1,2$ and real exponents $\beta_{\pm}$ for $l = 3,4$.

As an example of complex exponents, consider the $l=2$ case, $AdS_2 \times S^2$ slicing, with $a=5$ and complex exponents $\b_{\pm} = -2 \pm i$. We expect a spiral in the $(\theta_{(0)}, \theta_{(2)})$ plane. Figure \ref{d7ads2coeffs} shows the numerical result for $\theta_{(2)}$ as a function of $\theta_{(0)}$, which indeed exhibits a spiral, so we again have a first-order transition. In this case, the transition occurs at the critical value $\theta_{(0)}^{crit} = m / (2 \pi \alpha') = 1.6557$ or equivalently $m = \sqrt\lambda\, \theta_{(0)}^{crit}/(2\pi) = 0.2635 \, \sqrt\lambda $ times the inverse of the $AdS$ curvature radius.

\begin{FIGURE}
{
\includegraphics[width=0.48\textwidth]{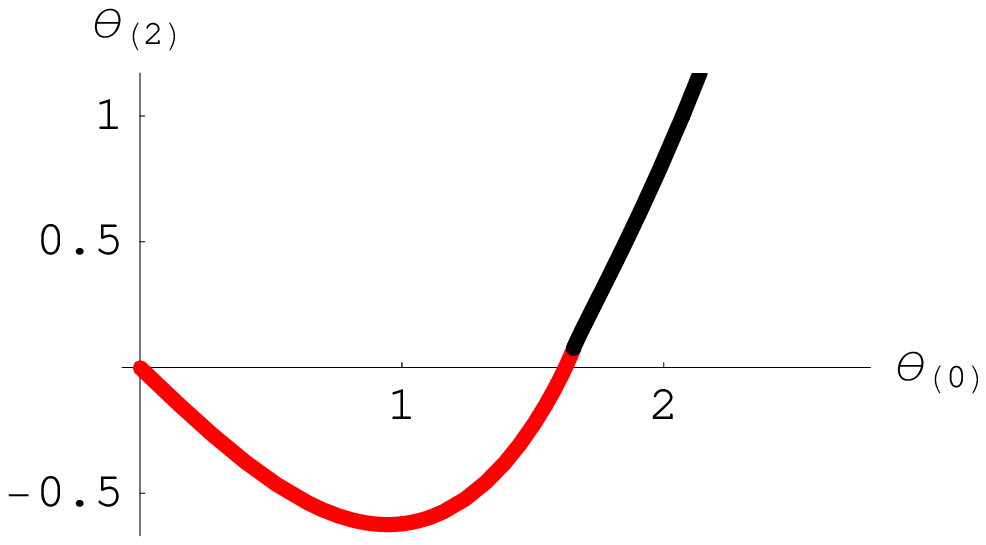} \hfil
\includegraphics[width=0.48\textwidth]{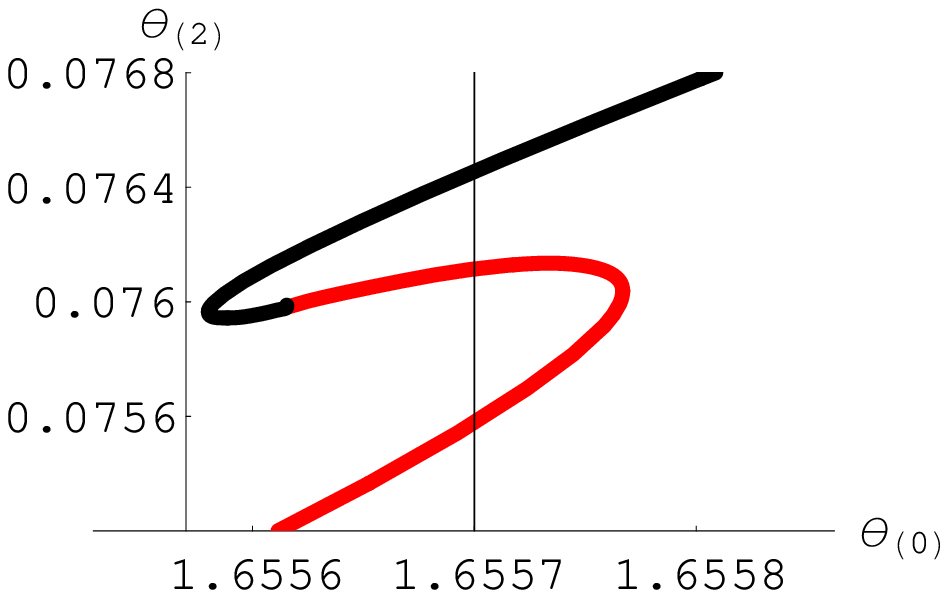}\\
(a) \hfil\hfil (b)
\caption{\label{d7ads2coeffs}(a.) $\theta_{(2)}$ as a function of $\theta_{(0)}$ for the D7-brane in $AdS_2 \times S^2$-sliced thermal $AdS_5$. (b.) Close-up of (a.) near the critical solution. The vertical line indicates where the transition occurs, at the critical value $\theta_{(0)}^{crit} \approx 1.6557$. }
}
\end{FIGURE}

We have two examples of real exponents. The first example is the $l=3$ case, $S^1 \times S^3$ slicing, which was examined above (see figures~\ref{d7action} and \ref{d7coeffs}) and will be examined in detail in the next section. The second example is the $l=4$ case, $S^4$ slicing, with $a = 7$, giving real exponents, $\b_{\pm} = -3 \pm \sqrt{2}$. We expect $\theta_{(2)}$ to be single-valued as a function of $\theta_{(0)}$. Figure \ref{d7s4coeffs} shows the numerical result for $\theta_{(2)}$ as a function of $\theta_{(0)}$, which is indeed single-valued. We have verified numerically that the solutions exhibit scaling behavior with the exponents $\b_{\pm} = -3 \pm \sqrt{2}$. In this case the critical solution has $\theta_{(0)}^* \approx 2.6395$ and $\theta_{(2)}^* \approx 14.5943$. In terms of SYM theory quantities the critical solution has $m^* = \sqrt{\lambda \,} \theta_{(0)}^*/(2\pi) \approx 0.42 \sqrt{\lambda}$ times the inverse of the AdS curvature radius.

\begin{FIGURE}
{
\includegraphics[width=0.48\textwidth]{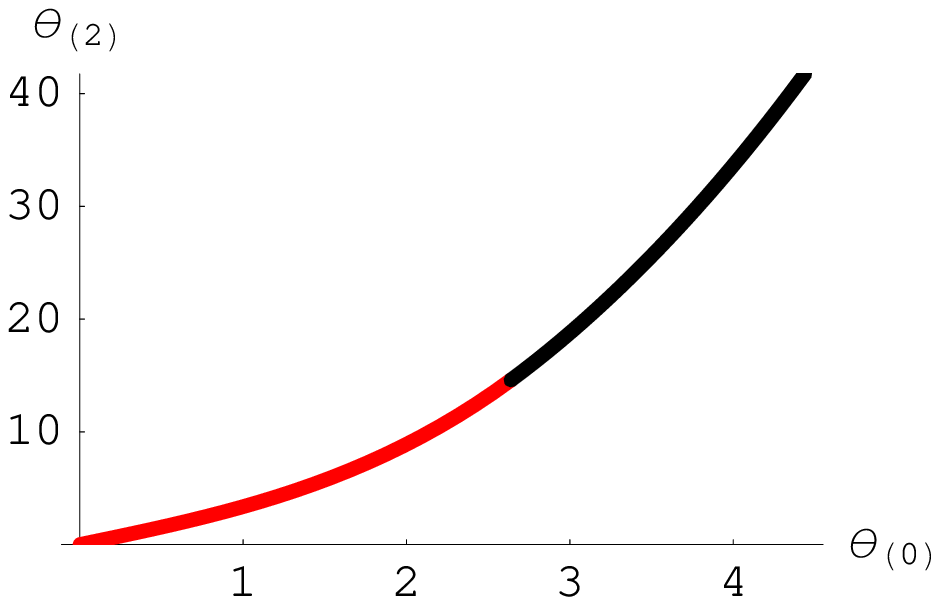} \hfil
\includegraphics[width=0.48\textwidth]{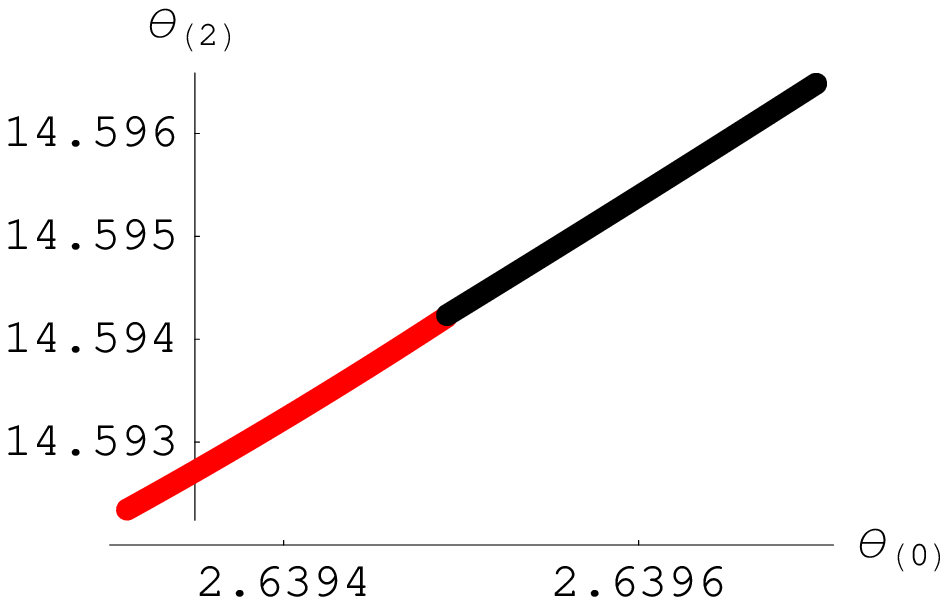}\\
(a) \hfil\hfil (b)
\caption{\label{d7s4coeffs}(a.) $\theta_{(2)}$ as a function of $\theta_{(0)}$ for the D7-brane in $S^4$-sliced thermal $AdS$. (b.) Close-up of (a.) near the critical solution. The critical solution (where the red and black curves meet) has $\theta_{(0)}^* \approx 2.6395$ and $\theta_{(2)}^* \approx 14.5943$. }
}
\end{FIGURE}

\section{Characterizing the Transition}  \label{transition}

Let us summarize our story so far. Whether a \Dp-brane probe exhibits a spiral in the $(\th_{(0)},\th_{(2\Delta -i + 1)})$ plane or not depends upon the value of $a$, the dimension of the submanifold of the critical embedding that collapses to zero volume. When $a < 6$, a spiral does appear and, as discussed above, we may conclude that the dual SYM theory has a first order phase transition associated with the flavor fields.

To study what happens when $a \geq 6$, we will restrict our attention to the D7-brane in $S^1 \times S^3$ slicing. The D7-brane undergoes a topology change, but $\theta_{(2)}$ is single-valued as a function of $\theta_{(0)}$ (or mass). In the SYM theory,  we expect a transition to occur, but evidently it is not first order.

In this section, we will exhibit divergences at the critical point in several observables in the dual SYM theory, and examine the associated critical exponents. We will also show that the transition manifests itself in the meson spectrum as a cusp in the spectrum of a particular scalar meson. We will also argue that the transition will be ``smoothed out'' for any large but finite value of $\lambda$, or in other words that the transition is an artifact of the $\lambda \rightarrow \infty$ limit.

\subsection{The Behavior of $\lang \Om \rang$}  \label{vev}

We want to know what the scaling transformation Eq.~(\ref{alphascaling}) implies for the observable $\lang \Om \rang$. Returning to Eq.~(\ref{coeffdiffs}), we perform the scaling transformation in Eq.~(\ref{alphascaling}), using $1 - \b_+ = 3$ and $1-\b_- = 4$, with the result

\begin{subequations}
\begin{equation}
\th_{(0)} - \th_{(0)}^* = A^+_{(0)} \, \a_+ \, \m^3 + A^-_{(0)} \, \a_- \, \m^4
\,,
\end{equation}
\begin{equation}
\th_{(2)} - \th_{(2)}^*= A^+_{(2)} \, \a_+ \, \m^3 + A^-_{(2)} \, \a_- \, \m^4
\,.
\end{equation}
\end{subequations}
Notice again that, as in Eq.~(\ref{coeffdiffs}), the form of these equations (and the equations below that follow from them) is the same for both $\th_{(0)} > \th_{(0)}^*$ and $\th_{(0)} < \th_{(0)}^*$. Using the first equation to eliminate $\mu$ gives, to leading order in the deviation from criticality, $\m \sim (\th_{(0)} - \th_{(0)}^*)^{1/3}$. Plugging this into the second equation will generate a non-analytic $(\th_{(0)} - \th_{(0)}^*)^{4/3}$ term as well as an analytic $(\th_{(0)} - \th_{(0)}^*)$ piece. Translating to field theory quantities using $\th_{(0)} \propto m$ and Eq.~(\ref{d7vev}), we find that $\lang \Om \rang$ contains non-analytic terms of the form
\begin{equation}
    \lang \Om \rang
    \sim (m - m^*)^{4/3} + \cdots \,,
\end{equation}
in addition to contributions analytic in $m$ near $m^*$. Here $\cdots$ is an expansion in higher powers of $(m - m^*)^{1/3}$. The key point is that $\lang \Om \rang $ has a divergent second derivative with respect to $m$ at $m^*$. The expectation $\lang \Om \rang $ is the derivative of the free energy $F$ with respect to $m$, and hence the \textit{third} derivative of the free energy diverges as $m \ra m^*$,
\begin{equation}
\frac{\partial^3}{\partial m^3} \, F \sim (m-m^*)^{-2/3} \,,
\end{equation}
from which we identify a critical exponent of $2/3$.

\subsection{Static Test Charges}  \label{wilson}

Consider, in the dual SYM theory, the interaction between static fundamental-representation test charges. Focus, for simplicity, on the case of an infinitely heavy quark and antiquark which are maximally separated --- sitting at opposite poles of the $S^3$. The change in free energy due to inserting the static quark and antiquark is given by ($-T$ times) the logarithm of the expectation value of the Polyakov loop (or Wilson line) correlator, with the two Polyakov loops sitting at antipodal poles in space.

Roughly speaking, in the dual gravitational description (in the $\Nc\to\infty$ and $\lambda\to\infty$ limits), this Polyakov loop correlator is given by the regularized minimal area of a string worldsheet with an $S^1 \times S^1$ boundary, where the $S^1$'s wrap the thermal circle and are maximally separated in the boundary spacetime. More precisely, the objects that have simple supergravity descriptions are the supersymmetric extensions of Wilson loops (``Maldacena loops''). These correspond to adding a line integral of a linear combination of the SYM scalars to the exponent of the Polyakov loop \cite{Maldacena:1998im}. Exactly what linear combination is determined by choosing, arbitrarily, some six-dimensional unit vector, or equivalently some point in $S^5$. In the dual description, the boundary of the string worldsheet must be located at the chosen point in the $S^5$ factor of the boundary geometry.

We will focus, for simplicity, on the maximally symmetric case where the Maldacena loops are located at a pole of the $S^5$, or in other words are maximally separated (in the boundary geometry) from the D7-brane wrapping an $S^3$ equator of the $S^5$. This choice preserves the full $SO(4)$ $R$-symmetry of the SYM theory with fundamental hypermultiplets.
The antipodal locations of the loops means that the correlator is also invariant under an $SO(3)$ subgroup of the $SO(4)$ spatial rotation symmetry group, as well as $U(1)$ time translations.

Given this setup, an extremal string worldsheet, preserving all the
symmetries of the correlator, will be one that stretches across
global $AdS_5$ while everywhere sitting at the pole of the $S^5$.
This is illustrated in figure \ref{soupcan} (B.).

\begin{FIGURE}
{\includegraphics[width=0.850\textwidth]{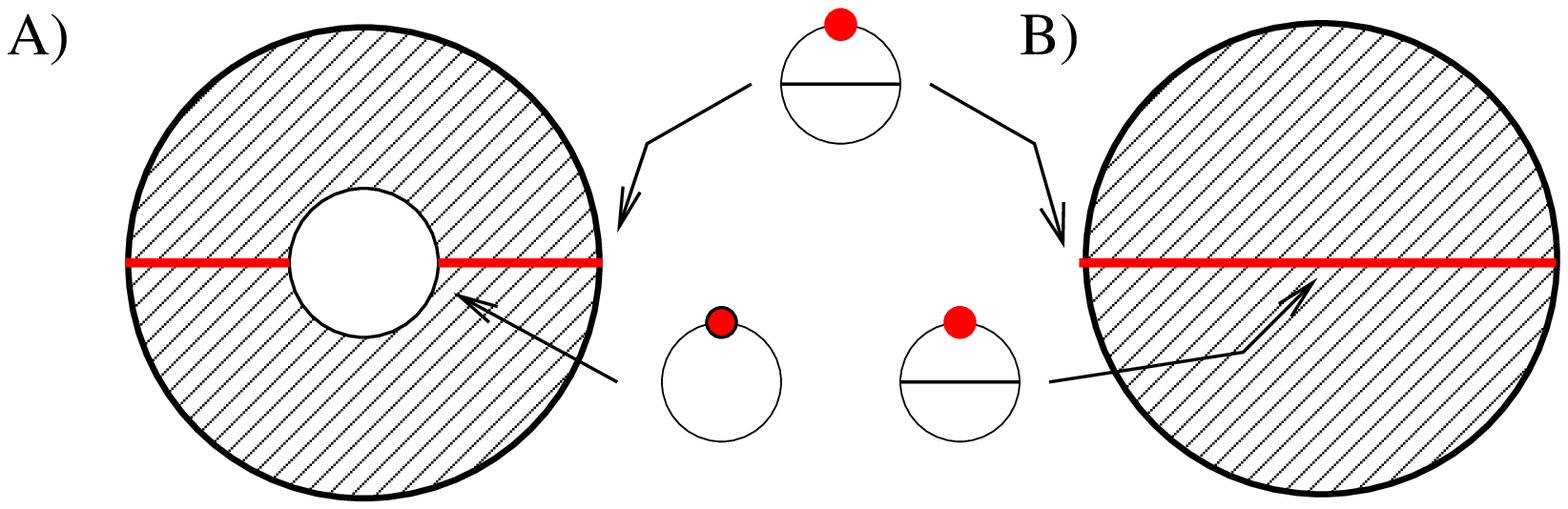}
\caption{\label{soupcan}(\textbf{A}.) The disk depicts global $AdS_5$ in $S^1 \times  S^3$ slicing. The perimeter represents the boundary while the center represents the center of $AdS_5$. The shaded/striped area represents a D7-brane that ends before reaching the center. The red lines represent two segments of a string, each stretching from the boundary to the D7-brane's endpoint. Each of the three circles between (A.) and (B.) represents the $S^5$. The D7-brane, wrapping an $S^3 \subset S^5$, is depicted as a black line. The location of the string(s) is depicted by a red dot. At the boundary, the D7-brane wraps the equatorial $S^3$, but at its endpoint it reaches the pole of the $S^5$, where the strings also sits (of the circles between (A.) and (B.), compare the top one and the lower left one). (\textbf{B}.) The same picture as in (A.), but now the D7-brane does not end at finite radial coordinate, rather it extends all the way to the center. Indeed, the figure depicts the trivial embedding, in which the D7-brane wraps the equatorial $S^3$ everywhere. Notice that in this case the string and D7-brane never coincide on the $S^5$.}}
\end{FIGURE}

If the D7-brane embedding fills the the whole $AdS_5$ space and
reaches the center, then the wrapped $S^3 \subset S^5$ never
collapses to zero volume (so $0 < \theta(\rho=0) < \frac{\pi}{2}$),
and the D7-brane never intersects the string worldsheet. This is not
the case for D7-brane embeddings that end outside the center (at
some $\rho = \bar\rho$ where $\theta(\bar{\rho}) = \frac{\pi}{2}$).
Such D7-branes reach the pole of the internal $S^5$ and intersect
the string worldsheet at $\bar\rho$. This allows the string to
break, so that its worldsheet ends on the D7-brane instead of
extending all the way to the center of global $AdS$. This is illustrated in figure \ref{soupcan} (A.). Phrased
differently, for Minkowski embeddings of the D7-brane, a
maximal-symmetry string worldsheet with an $(S_1)^4$ boundary exists
--- with two of the $S^1$'s on the $AdS$ boundary and the other two on
the D7-brane at a point of closest approach to the global $AdS$
center. Again, the two possibilities are depicted in figure \ref{soupcan}.

The potential energy of the static quark and antiquark is directly proportional to the length of the string.
The difference in length of the strings in the two configurations is just the diameter of the ``hole'' in the center of the D7-brane configuration. For the near-critical embeddings, this is directly proportional to the scaling parameter $\m$.
Therefore the difference in the static quark-antiquark potential energy (for this maximally-symmetric configuration)
between near-critical and critical values of the quark mass will proportional to $\mu \sim (m-m^*)^{1/3}$ (or $(m^*-m)^{1/3}$ for $m<m^*$)--- and thus
non-analytic in the quark mass at $m^*$.

The behavior of this Polyakov loop correlator has an intuitive explanation in terms of screening lengths. To understand this, first consider the physics of QCD (in flat space) at low temperature, where two infinitely massive test quarks will have a flux tube, or QCD string, stretching between them. The energy $E$ in this QCD string is its tension, $\sigma$, times its length, $L$: $E = \sigma L$. The theory also has dynamical quarks, however, which can cause sufficiently long flux tubes to break. If the test quarks are pulled sufficiently far apart, then the energy in the string will exceed the point at which the creation of a dynamical quark-antiquark pair is energetically favorable. In other words, when $E > 2 m$ for a dynamical quark mass $m$, the string can snap and the single heavy-heavy meson will decay into two heavy-light mesons. Define a ``screening length'' as the length $L_s$ at which the pair production occurs and the QCD string snaps. Equating $\sigma L_s = 2 m$, we see that the screening length is \textit{proportional} to the mass of the quark, $L_s \propto m$.

Now consider the $\Nfour$ SYM theory formulated on $\mathbb{R}^3$ coupled to massive $\Ntwo$ hypermultiplets. The story is a bit different since the theory is conformal. This implies that the static quark potential is Coulombic, and vanishes at large separation. Na\"{i}vely, one might think that this would imply that pair production of dynamical massive quarks would never be energetically favorable. (After all, hydrogen is not unstable to electron/positron pair creation!) This argument relies on our \textit{weak}-coupling intuition, however. If the coupling is sufficiently strong, then the light-heavy binding energy can become comparable to the light quark mass $m$. In this regime a heavy-heavy meson, pulled apart to large separation, can indeed decay to a pair of light-heavy mesons via pair production of a light quark-antiquark pair.
This scenario is sometimes called Gribov confinement \cite{Gribov}. For the $\Nfour$ SYM theory, this process was analyzed in detail in Ref.~\cite{Karch:2002xe}. The screening length, beyond which the system will pair produce dynamical quarks, must by scale invariance be proportional to the inverse of the light quark mass, $L_s \propto 1/m$.

Finally, consider the $\Nfour$ SYM theory formulated on $S^3$ with $\Ntwo$ matter. As we lower the dynamical quark mass we find that at some point this screening length becomes larger than the diameter of the sphere, $L_s > 2R_3$. Beyond this point, a (maximally separated) static quark-antiquark pair is stable against pair production of dynamical quarks. Our Polyakov loop correlator exhibits precisely this behavior.

\subsection{The Meson Spectrum}  \label{meson}

In the $\Nfour$ SYM theory coupled to $\Ntwo$ matter on $S^1 \times \R^3$, the first-order phase transition of the fundamental-representation fields is characterized by ``meson melting.'' At low temperature the meson spectrum is gapped and discrete while at high temperature it becomes gapless and continuous \cite{Mateos:2006nu,Hoyos:2006gb,Mateos:2007vn}. A
natural question for the theory on $S^3$, then, is what happens to the meson spectrum at the transition we have found?

For conceptual simplicity, and for technical reasons that will become clear shortly, in this section only we will take the temperature $T$ to be precisely zero. We are allowed to do so because, as explained above, the D7-brane physics we are interested in is independent of temperature in the low-temperature phase. We are thus free to take $T = 0$ and study the SYM theory formulated on the four-manifold $\R \times S^3$ with $\R$ the (Minkowski-signature, non-compact) time direction.

At zero temperature we have a simple intuitive picture of what happens. Consider first the infinite-volume case at finite temperature. If the mass $m$ is fixed and we heat the system up, stable bound mesons fall apart at the ``melting'' temperature. What we are instead doing here may be viewed as fixing the mass and then squeezing the system,
at zero temperature, into smaller and smaller volume and asking what happens to mesonic bound states as $R_3$ passes through the critical radius $R_3^* = 0.1748 \sqrt{\lambda} / m$. We expect the mesons to fall apart as their zero
point energy due to confinement within the finite volume becomes larger than their binding energy.

Before discussing the meson spectrum in detail, we should define precisely what we mean by a meson mass.  On $\R \times S^3$, a state is classified by its $SO(4)$ angular momentum $l(l+2)$, for non-negative integer $l$, and its energy eigenvalue $\o$, governing the behavior under time translations. We will use this energy eigenvalue $\o$ to characterize a meson state, and will refer to $\o$ as the meson mass.\footnote{In some literature on holographic mesons in flat space, the name ``meson mass'' is used for the magnitude of an imaginary spatial wave vector $\vec{k}$ for which an eigenstate of the supergravity small fluctuation operator exists. These eigenvalues characterize the leading long distance fall-off of equilibrium Euclidean space correlators in the dual field theory. On $S^3$, no (interesting) analogue of this definition exists. The spatial (angular) momentum simply takes values $l(l+2)$ for non-negative integers $l$; since the space is finite we have no notion of the asymptotic behavior of correlation functions.} We will only consider states that are in an $l=0$ $s$-wave on $S^3$, and compute the dependence of $\o^2$ on the quark mass $m$, or more precisely (but equivalently) on $\theta_{(0)}$. States with higher $l$ are expected to have higher energy. As explained in Ref.~\cite{Kruczenski:2003be}, the meson spectrum can be computed in the supergravity description by expanding the DBI action to quadratic order in fluctuations and then solving the resulting linearized equations of motion. The fluctuations may be those of either the embedding geometry or of the D7-brane worldvolume gauge fields.

\begin{FIGURE}[t]
{
\includegraphics[width=0.48\textwidth]{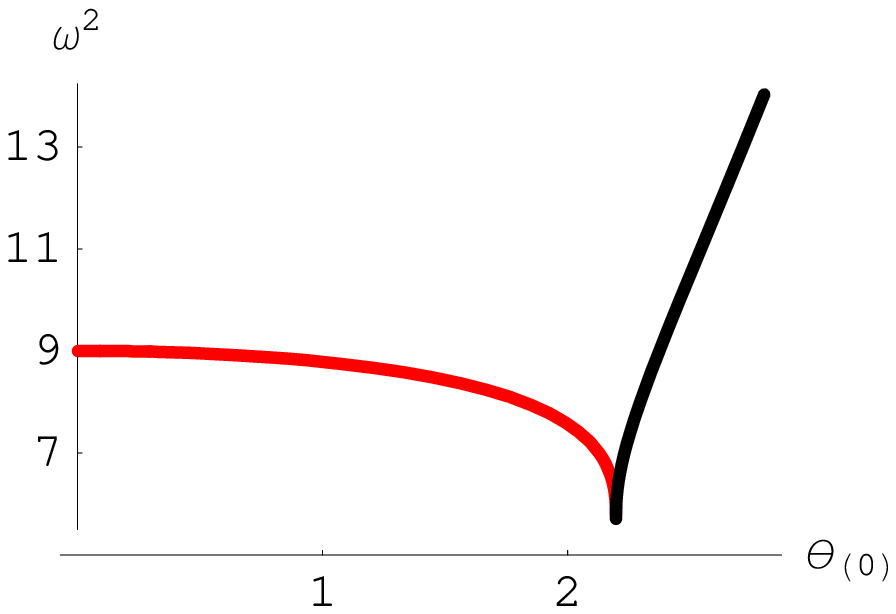} \hfil
\includegraphics[width=0.48\textwidth]{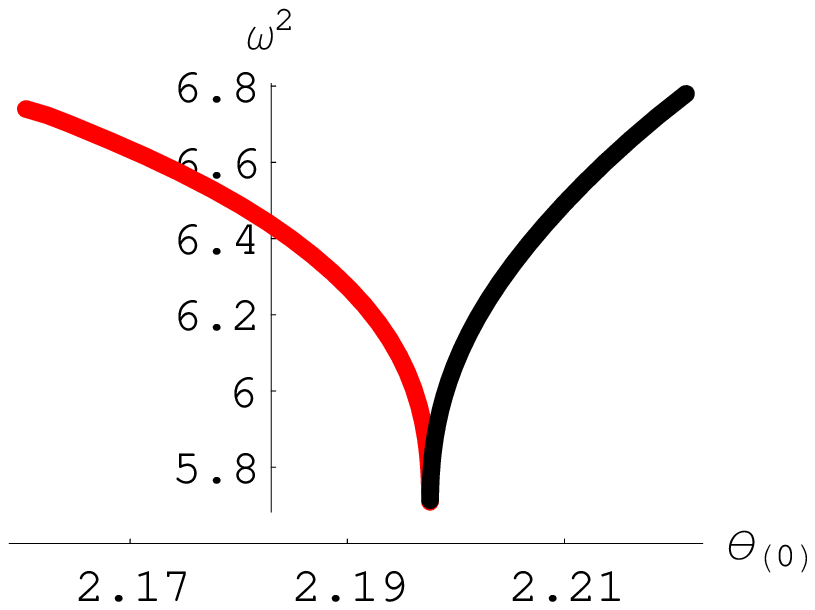}\\
(a) \hfil\hfil (b)
\caption{\label{spectrum}(a.) Scalar meson frequency squared (times the $AdS$ radius squared) versus $\theta_{(0)}$. The critical solution has $\theta_{(0)}^* = 2.198$ and $(\o^*)^2 = 5.7$. (b.) Close-up of (a.) near the critical solution. Our analytic arguments (confirmed numerically in figure \ref{onethird}) show that the curves behave as $(\theta_{(0)} - \theta_{(0)}^*)^{1/3}$ near the critical solution.}
}
\end{FIGURE}

We first study a fluctuation of the D7-brane geometry of the form
\begin{equation}
\th(\rho) \ra \th(\rho,t) = \th_B(\rho) + \phi(\rho) e^{i \o t} \,,
\label{fluc}
\end{equation}
where $\th_B(\rho)$ is a static background solution, as computed in section \ref{low}. This fluctuation will be dual to a scalar meson in the SYM theory. To determine $\o$, we employ a shooting technique in which we fix the boundary condition for $\phi(\rho)$ at the $AdS$ boundary and iteratively adjust $\o$ until the physical boundary condition at the D7-brane endpoint (or the center) is satisfied. At the $AdS$ boundary, we demand that the fluctuation be normalizable and hence must scale, in Fefferman-Graham coordinates, as $O(\e^3)$ with the cutoff $z=\e$. For D7-branes that reach the center, we then demand that the fluctuation must have vanishing first derivative at the center. For D7-branes ending away from the center, we require that the $S^3 \subset S^5$ collapse without a conical deficit, as explained in section \ref{probe}. Translating the results into field theory quantities, the resulting spectrum of $\o^2$ versus $\theta_{(0)} = 4 \pi \sqrt{\lambda} \, m R_3$ for this scalar meson is shown in figure \ref{spectrum}. We see a pronounced kink precisely at the critical solution, $\theta_{(0)} = \theta_{(0)}^*$. The kink does not extend down to $\omega^2 = 0$; the minimal value is $\omega^2 \approx 5.7$ times the square of the $AdS$ radius, or equivalently $\o^2 \approx 22.8 R_3^2$.

Using scaling symmetry arguments, we can show analytically that this kink in the scalar meson spectrum has the precise form $\o - \o^* \sim (m-m^*)^{1/3}$ as $m \to m^*$ (from above or below). We begin by solving analytically for the fluctuation in the near-center limit. At zero temperature the action for $\th(\r,t)$ is
\begin{equation}
\tilde{S}_{D7} = \N_{D7} \int d\r \> dt \> (\cos \th)^3 \, (\sinh\r)^3 \sqrt{(1+ \th'^{\,2})(\cosh^2 \r - \dot{\th}^2) + \dot{\th}^2 \, \th^{'2}} \,.
\end{equation}
We take the same near-center limit as in section \ref{low}, insert Eq.~(\ref{fluc}), and expand to quadratic order in the fluctuation $\phi(\r)$ to find the linearized equation of motion. We choose the background solution to be the critical solution $\theta_B(\rho) = \th^*(\r) = \frac{\pi}{2} + \r$, so the frequency of the fluctuation is the critical one, $\o = \o^*$. The resulting equation for $\phi(\r)$,
\begin{equation}
\r^2 \, \phi''(\r) + 6 \r \, \phi'(\r) + (6 + 2 \o^{*2} \r^2) \, \phi(\r) = 0,
\label{fluceq}
\end{equation}
has a solution
\bea
\phi(\r) & = &
    \frac{1}{\r^3}
    \left [
    c_0^* \> \cos \bigl( \sqrt{2} \o^* \r \bigr)
    + c_1^* \> \frac{\sin \bigl( \sqrt{2} \, \o^* \r \bigr)}{\sqrt 2\, \o^*}
    \right ]
\\ & = &
    c_0^* \, \r^{-3} +
    c_1^* \, \r^{-2} +
    O(\rho^{-1}) \,,
\nonumber
\label{flucsol}
\eea
where $c_0^*$ and $c_1^*$ are integration constants. The background is the critical solution, for which the brane ends at the center, $\r = 0$. Normalizability at $\r=0$ requires $c_0^* = 0$. We could then fix the value of $\o^*$ by imposing normalizability at the $AdS$ boundary as follows.

Normalizability at the boundary requires that $\phi(\r)$'s contribution to the leading, non-normalizable asymptotic
coefficient (which was $\th_{(0)}$ in Fefferman-Graham coordinates) must vanish. We denote this contribution as
\begin{equation}
\Phi^*(c_1^*,\o^*) = f_1(\o^*) \, c_1^* + O(c_1^{*2}) = 0 \,,
\end{equation}
where, as indicated, this may be a function of $c_1^*$ and $\o^*$ and we have linearized in $c_1^*$, which is taken to be small. We could then solve the equation $f_1(\o^*)=0$ to find the value of $\o^*$.

We next want to find the shifted frequency, $\o = \o^* + \d \o$, of fluctuations of near-critical solutions. More precisely, we need to determine how $\delta \o = \o - \o^*$ scales with $\mu \sim (m-m^*)^{1/3}$. Near-critical solutions will have nonzero values for $c_0$ and $c_1 = c_1^* + \d c_1$ so that:
\bea
\label{nearcriticalflucequation}
\th(\r,t) & = & \r
+ \a_- \, \r^{-3}
+ \a_+ \, \r^{-2}
+ \phi(\r) \, e^{i \o t}
\\ & = &
\r
+ (\a_- + c_0 \, e^{i \o t}) \, \r^{-3}
+ (\a_+ + c_1 \, e^{i \o t}) \, \r^{-2}
+ O(\rho^{-1})  \,.
\nonumber
\eea
We see that $c_0$ must scale the same way as $\a_-$, and $c_1$ the same as $\a_+$, that is $c_0 \ra \m^4 \, c_0$
and $c_1 \ra \m^3 \, c_1$. To fix the value of $\o$, we again impose the condition of normalizability at the $AdS$ boundary, which requires that the coefficient $\Phi(c_1,c_0,\o)$ of the non-normalizable term vanish,
\bea
0 = \Phi(c_1,c_0,\o) & = & f_1(\o) \, c_1 + f_2(\o) \, c_0
\\ & = & \left [ f_1(\o^*) + f_1'(\o^*) \d \o \right]
\left ( c_1^* + \d c_1 \right )
+ \left[ f_2(\o^*) + f_2'(\o^*) \, \d \o \right ] c_0 \nonumber
\\ & = & f_1'(\o^*) \, \d \o c_1^* + f_2(\o^*) \, c_0  \,,\nonumber
\eea
where we have linearized everything treating $\d c_1$, $c_0$ and $\d \o$ as the same order of smallness and used $f_1(\o^*) = 0$. We may immediately solve for $\d \o = \o - \o^*$ with the result
\begin{equation}
\o - \o^* = - \frac{f_2(\o^*)}{f_1'(\o^*)} \, \frac{c_0}{c_1^*} \,.
\label{eq:domega}
\end{equation}
Notice $f_2(\o^*)$ and $f_1'(\o^*)$ do not transform under the scaling symmetry: they are just numbers. Since $c_0$ scales as $\mu^4$ while $c_1^*$ scales as $\mu^3$, the result of Eq.~(\ref {eq:domega}) shows that $\o - \o^* \sim \m \sim (m-m^*)^{1/3}$. Notice that these arguments hold for both $m > m^*$ or $m<m^*$ (the results do not depend on whether the near-critical solution in Eq.~(\ref{nearcriticalflucequation}) has $\theta_{(0)}>\theta_{(0)}^*$ or $\theta_{(0)}<\theta_{(0)}^*$). Figure \ref{onethird} shows our numerical data for $\ln |\o - \o^*|$ versus $\ln |\theta_{(0)} - \theta_{(0)}^*|$ and a linear fit to the data, with good agreement between this asymptotic form and the numerical results.

\begin{FIGURE}[t]
{
\includegraphics[width=0.47\textwidth]{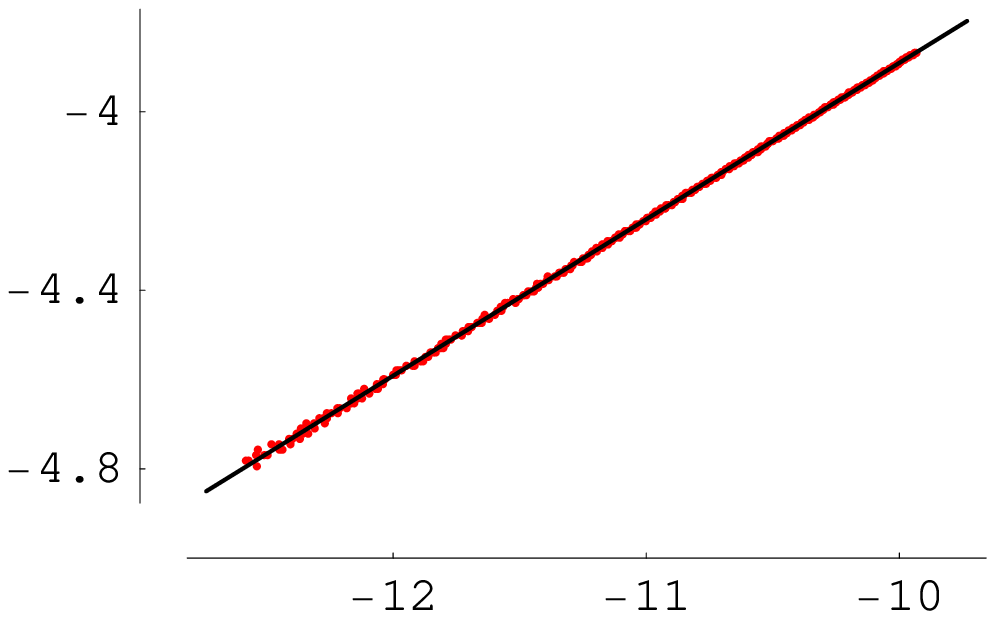} \hfil
\includegraphics[width=0.47\textwidth]{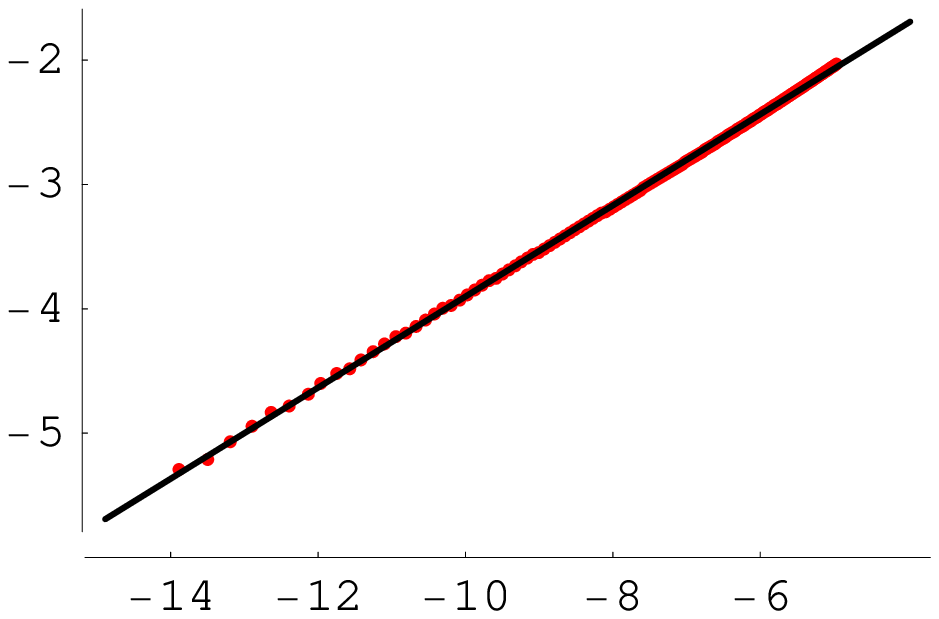}\\
(a) \hfil\hfil (b)
\caption{\label{onethird}(a.) $\ln | \o - \o^*|$ (with $\o$ in units of the inverse $AdS$ radius) versus $\ln|\theta_{(0)} - \theta_{(0)}^*|$ for D7-branes that end at the center, which have $\theta_{(0)} < \theta_{(0)}^*$ (corresponding to the red curves in the previous figure). The solid black line is a numerical fit to a functional form $C_1 + C_2 \ln|\theta_{(0)} - \theta_{(0)}^*|$. Our analytic argument predicts $C_2 = 1/3$. The numerical result is $C_1 \approx -0.38$ and $C_2 \approx 0.35$. (b.) The same quantities as in (a.) but now for D7-branes that end away from the center, which have $\theta_{(0)} > \theta_{(0)}^*$ (the black curves in the previous figure). The solid black line is a numerical fit of the same form as in (a.) with the result $C_1 \approx -0.24$ and $C_2 \approx 0.36$.}
}
\end{FIGURE}

\begin{FIGURE}[t]
{
\centering
\includegraphics[width=0.58\textwidth]{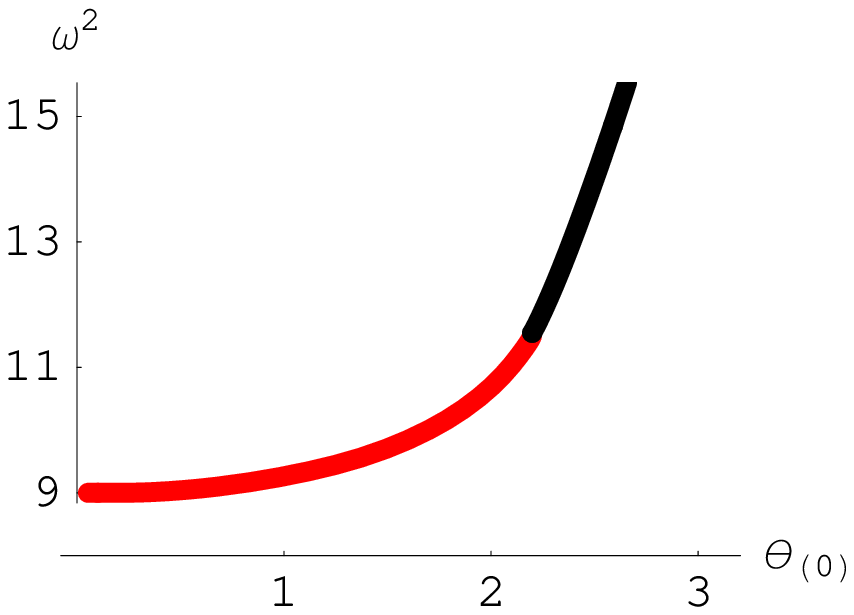}
\caption{\label{otherscalar} Scalar meson frequency squared (times the $AdS$ radius squared) versus $\theta_{(0)}$ for the meson holographically dual to the second fluctuation of the D7-brane geometry. The punchline here is that no ``kink'' appears, in contrast to the spectrum in figure \ref{spectrum}.}
}
\end{FIGURE}

This non-analytic behavior appears to manifest itself only for the geometric fluctuation of the embedding function $\th(\r)$. Solving for the other fluctuation in the geometry (the other $S^5$ direction orthogonal to the $S^3$) we can compute the spectrum for a second scalar meson. The result appears in figure \ref{otherscalar}. We see no kink at $\theta_{(0)}^*$. We also computed the meson spectra corresponding to fluctuations of the D7-brane's worldvolume gauge field as described in Ref.~\cite{Kruczenski:2003be}. These meson spectra are similar to figure \ref{otherscalar}. In particular, they exhibit no kinks. We confirmed that all of our meson spectra reduce to the known results of Ref.~\cite{Kruczenski:2003be} in the limits of either zero or large mass.

\subsection{Light String States}  \label{stringstates}

In section \ref{vev}, we showed that the third derivative of the free energy $F$ diverges at $m^*$. In other words, some three-point coupling in an effective theory describing flavored mesons in this strongly-coupled SYM theory is diverging as $m$ approaches $m^*$. One question this immediately raises is whether this power-law growth of the three-point coupling continues for arbitrarily small values of $m-m^*$, and if so, what is the physical meaning of this singularity? We will argue that for any large, but \textit{finite}, values of $\lambda$, the scaling regime will be cut off at small values of $m-m^*$, so the divergence in the three-point coupling is an artifact of the strict $\lambda\to\infty$ limit.

Our evidence comes from the string theory side of the correspondence: the scaling regime is cut off by stringy corrections. To see this, note that the scalar curvature of the D7-brane's induced metric, ${\cal R}_{D7}$, in the near-center limit, is
\begin{equation}
{\cal R}_{D7} = -12  \, \frac{\d\theta^2 \, \d\theta'^{\,2} -\frac 32\, \rho \, \d\theta \, \d\theta' + \rho^2}{\rho^2 \, \d\theta^2 \, (1+\d\theta'^{\,2})} \,.
\end{equation}
For the critical embedding, $\d\theta^*(\r) = \r$, the curvature diverges as $\rho^{-2}$ near the endpoint $\rho=0$. For near-critical solutions, the curvature is finite at the endpoint (or center), but grows without limit as one approaches the critical solution\footnote
    {
    For the D7-brane in $S^1 \times \R^3$ slicing (with its first order transition),
    the scalar curvature also diverges for the critical solution,
    but in this case the critical solution lies on an unphysical
    (infinitely unstable) branch.
    }. For branes that reach the center, $\d\theta'=0$ at the center and hence ${\cal R}_{D7}$ scales as $\d\theta^{-2}$. For solutions ending away from the center, $\d\theta'$ diverges, and the curvature at the endpoint scales as $\rho^{-2}$. Under a scaling transformation, both $\rho$ and $\d\theta$ scale as $\mu$, so the curvature in either case will scale as $\mu^{-2}$.

When the scalar curvature of the induced metric becomes of order of the string scale, $\alpha'^{-1} \equiv \ell_s^{-2}$, stringy corrections to the DBI action will no longer be negligible, and hence our analysis of the scaling behavior (based entirely on the DBI action) will cease to be valid. In other words, higher order corrections will become important when $\m^{-2} \sim \ell_s^{-2}$. These corrections are due to excited open string modes whose mass na\"{i}vely goes as $\ell_s^{-1}$, but whose mass is reduced by a power of $\m$ when the endpoints of the string are in the high-curvature region. To illustrate this issue graphically, we plot $\ln[ -{\cal R}_{D7}]$ in figure \ref{s3curv}, evaluated at the point of closest approach to the center, as a function of the asymptotic coefficient $\th_{(0)}$.

\begin{FIGURE}[t]
{
\centering
\includegraphics[width=0.6\textwidth]{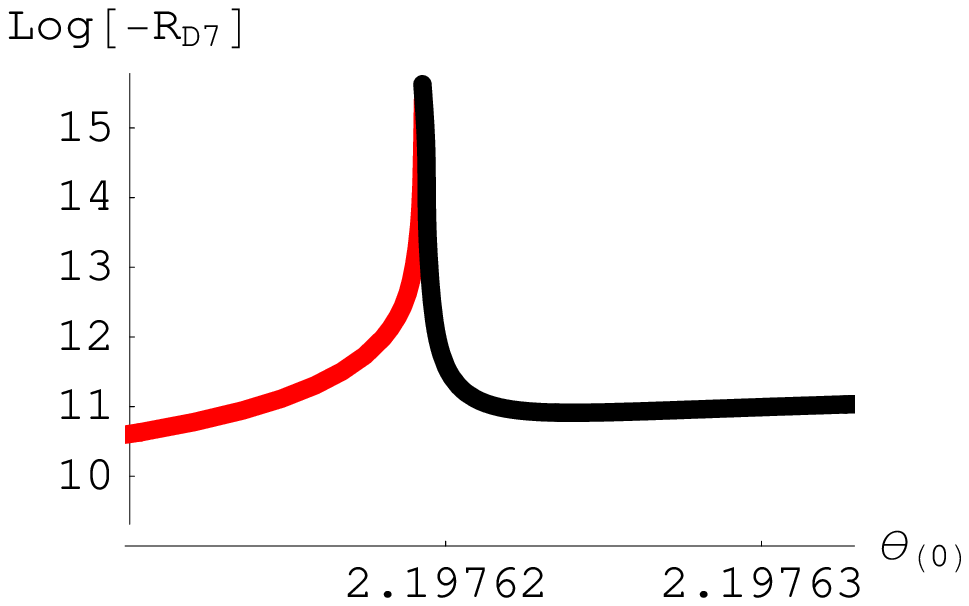}
\caption{\label{s3curv} Log of minus the scalar curvature of the D7-brane, $\ln [ -{\cal R}_{D7}]$ (in units of the $AdS$ radius), evaluated at the point of closest approach to the center of $AdS$, versus asymptotic coefficient $\th_{(0)}$ in $S^1 \times S^3$ slicing. The red curve represents D7-branes that reach the center, for which $\ln [ - {\cal R}_{D7}]$ is evaluated at the center. The black curve represents D7-branes that end away from the center, for which $\ln [ - {\cal R}_{D7}]$ is evaluated at the endpoint.}
}
\end{FIGURE}

Using $\a'^{-1} \equiv \ell_s^{-2} \sim \sqrt{\lam}$ and $\m \sim (\th_{(0)} - \th_{(0)}^*)^{1/3} = (m-m^*)^{1/3}$, the condition $\m^{-2} \sim \ell_s^{-2}$ becomes $(m-m^*) \sim \lambda^{-3/4}$. The properties of the boundary theory are thus governed by the critical exponents we calculated above, but only in the range of masses $\lambda^{-3/4}/R_3 \ll (m-m^*) \ll 1/R_3$. (Once again, the same relations are true for $m<m^*$, with $m-m^*$ replaced by $m^* - m$, etc.) In particular, in this window the three-point function of the zero-momentum $\Om$ operator grows as $(m-m^*)^{-2/3}$. This power law growth will eventually be cut off at the lower end of the scaling window. As $\lambda$ is a free parameter, however, we are free to consider the regime where this scaling window extends over arbitrarily many decades.

\section{Conclusion}  \label{conclusion}

The scaling symmetry of near-center or near-horizon probe \Dp-brane solutions in global $AdS$ has allowed us to determine when phase transitions associated with fundamental-representation fields coupled to $\Nfour$ SYM theory will be first order or continuous. When they are continuous, we find that the approach to criticality is governed by non-trivial critical exponents which can be calculated analytically using the supergravity description. We emphasize that the phase transitions we have found are finite-volume, large-$\Nc$, and large-$\lambda$ effects. The continuous transition we find in the D7-brane case can be interpreted as a meson binding/unbinding transition similar to the finite-temperature case analyzed in previous studies.

A finite volume in the flavored SYM theory can give rise to interesting new effects in the phase diagram, such as the appearance of our continuous phase transition. We have only explored a small part of the multi-dimensional phase diagram of this theory. Probe \Dp-brane techniques can be used to study systems at finite density \cite{Kobayashi:2006sb,Ghoroku:2007re,Karch:2007br,Mateos:2007vc,Faulkner:2008hm} or in background electric and magnetic fields \cite{Filev:2007gb,Filev:2007qu,Albash:2007bk,Erdmenger:2007bn,Albash:2007bq,Filev:2008xt,Filev:2009xp}. We expect a rich phase structure to emerge for these systems when confined to finite volume.

\section*{Acknowledgments}

This work was supported in part by the U.S. Department of Energy under Grant No.~DE-FG02-96ER40956. The work of A.O'B. was also supported in part by the Jack Kent Cooke Foundation and by the Cluster of Excellence for Fundamental Physics ``Origin and Structure of the Universe.'' A. O'B. would like to thank the Perimeter Institute and the Aspen Center for Physics for hospitality during the completion of this work.

\bibliography{biblio}
\bibliographystyle{JHEP}

\end{document}